\documentclass[aps,prb,twocolumn,nopacs,amssymb]{revtex4-1}

\usepackage[dvips]{graphicx}
\usepackage{dcolumn}
\usepackage{bm}
\usepackage{color}
\def\be{\begin{equation}}
\def\en{\end{equation}}
\def\bea{\begin{eqnarray}}
\def\ena{\end{eqnarray}}
\def\bec{\begin{equation}\begin{array}{rcl}}
\renewcommand{\theequation}{\arabic{section}.\arabic{equation}}

\def\p{\partial}

\def\gs{\gtrsim}
\def\ls{\lesssim}

\newcommand{\av}[1]{\langle{#1}\rangle}

\newcommand{\bi}[1]{\mbox{\boldmath$#1$}}
\newcommand{\pp}[2]{\frac{\partial {#1}}{\partial {#2}}}

\def\aw{\stackrel{\leftrightarrow}{w}}
\def\aI{\stackrel{\leftrightarrow}{I}}
\def\a1{\stackrel{\leftrightarrow}{1}}

\begin{document}
\title{Applying  electric field 
to  charged and polar particles between 
 metallic plates:  Extension of the Ewald method}  
\author{Kyohei Takae}
\author{Akira Onuki}
\affiliation{Department of Physics, Kyoto University, Kyoto 606-8502, Japan}
\date{\today}


\begin{abstract} 
We develop an efficient   Ewald method 
of molecular dynamics simulation 
for calculating  the electrostatic interactions 
among charged and polar particles between parallel metallic plates, 
where  we may apply an electric field with an arbitrary size. 
We use the fact  that the potential 
from the surface charges is  equivalent to the sum of 
those from  image charges and  dipoles located 
outside the cell.  We present  simulation  results 
on  boundary effects  of charged and polar fluids,  
formation of  ionic crystals, 
and formation of dipole chains, where the 
 applied  field and 
the image interaction are  crucial. For  polar fluids,
we find a large  deviation of   the 
classical Lorentz-field relation 
between  the local field and the applied field due to 
pair correlations along the applied field.  
As  general aspects,  we clarify 
the difference between the potential-fixed and 
the charge-fixed boundary conditions  
and examine  the relationship between 
 the discrete particle 
description  and the 
 continuum electrostatics.
\end{abstract}


\maketitle



\section{Introduction}

In many problems in physics and chemistry, 
the electrostatic potential and field 
acting on each   constituting particle  need to be 
calculated \cite{Is}.  A large number of 
 simulations have been performed to 
accurately estimate  the long-range 
 electrostatic  interactions among 
 charged and polar particles \cite{Allen,Frenkel}. 
The Ewald method is a famous technique 
 for efficiently summing these   
interactions using  the Fourier transformation. 
It  was originally devised  for ionic crystals \cite{Ewald} 
and has been widely  used to investigate 
bulk properties of  charged and polar particles 
under the periodic boundary  condition in three dimensions (3D) 
\cite{Leeuw,Leeuwreview,Weis,Mazars}. It has also been 
modified  for filmlike systems  bounded by 
non-polarizable and insulating regions   
 under  the periodic boundary condition in the lateral directions 
\cite{Frenkel,Parry,Smith,Crozier,Yeh,KlappJCP,Tyagi,Smith1}.

However,  the Ewald method 
has not yet been successful 
  when    charged or  polar particles 
 are in contact with metallic or polarizable 
plates and when electric field 
is applied from outside.  Such situations  are ubiquitous  
in solids and soft matters.  
 In idealized metallic plates,  
the surface charges spontaneously appear such that the 
electrostatic potential is homogeneous within the plates, 
thus providing the well-defined 
boundary condition for the potential 
within the cell \cite{Landau}.  
Between two  parallel plates, we may 
control the  potential difference to apply an electric field.  
However, the electrostatic interaction between 
the surface charges and  the particles within the cell 
is highly nontrivial. 
On the other hand, such surface changes  are nonexistent for 
 the magnetic interaction.

Each charged particle  between parallel metallic plates 
induces  surface charges producing  a potential 
equivalent to   that  from  an infinite number of 
image charges   outside the cell. 
If a charged particle  approaches a metal wall, 
it is attracted by  the wall or by its nearest image. Some  mathematical 
formulae including these image charges in 
the Ewald sum    were presented   
by Hautman {\it et al}.\cite{Hautman} and by 
Perram and Ratner \cite{Perram}.   In the same manner, for each dipole  
in the cell, an infinite number of image dipoles appear.  A  
  dipole close to a metal surface  is  
attracted and aligned by its nearest image.   
 Accounting for these  image  dipoles,  
 Klapp \cite{Klapp}   performed  
Monte Carlo simulations of   500 dipoles  
interacting  with  the soft-core potential, 
but without applied electric field,   
 to find   wall-induced  ordering. 
The image effect is also  relevant 
in electrorheological fluids between metallic plates \cite{Hasley,Tao}. 

Inclusion of the image effect  in molecular dynamics  simulations 
of molecules near a conducting or polarizable 
surface is still  challenging  in 
a variety of important systems 
including proteins \cite{Corni}, 
polyelectrolytes \cite{Messina}, and colloidal 
particles \cite{colloid}.
Moreover, the effects of applied electric field 
remain largely unexplored  on the microscopic  level,  
while the   continuum electrostatics is  
well established \cite{Landau,OnukiNATO}.
For example, the local electric field acting on 
a dipole is known to be different from the applied electric field 
in dielectrics and polar fluids \cite{Onsager,Kirk},  
where  the difference is enlarged in  
highly polarizable systems.  
 In  contrast, a number of microscopic simulations 
 have been performed on  the effect of uniform 
 magnetic field  for systems of  magnetic dipoles 
\cite{magnetic,Weis,Mazars}. 

In this paper, we  hence aim 
to develop an efficient   Ewald method 
to treat charged and polar particles between parallel 
 metallic plates accounting for the image effect, 
where we apply  an electric field with an arbitrary size. In  the Ewald sum  
in this case, we can sum up  the terms 
homogeneous in the lateral $xy$ plane 
but inhomogeneous along the normal $z$ axis  
  into  a simple form and can  
calculated them precisely. This {\it one-dimensional} 
part of the electrostatic energy yields 
{\it one-dimensional} (laterally 
averaged)   electric 
field along the $z$ axis for each particle.  
 
Using our scheme, we present  some numerical results 
  under applied electric field,  
 including  the soft-core pair interaction and the 
wall-particle repulsive interaction. 
We confirm that accumulation of 
 charges and dipoles near metallic  walls 
gives rise to a  uniaxially symmetric, homogeneous interior  region. 
We  also examine  formation of ionic crystals \cite{super} and  
that of dipole chains \cite{Hasley,Tao,PG} 
under electric field. 
Here,  we are interested in 
the mechanisms of dipole   alignment   
near  metallic walls, 
which are caused  by  the image interaction 
or by the applied electric field. 
 We shall also  see that  
  the classical local  field 
relation  \cite{Onsager,Kirk} is  much violated 
in our dipole systems because of strong pair  correlations 
 along the applied  field.

The organization of this paper is as follows. 
In Sec.II,   we will 
extend the Ewald scheme  for charged particles 
between metallic plates.  
In Sec.III, we will further develop 
the Ewald scheme for point-like dipoles 
between metallic plates. 
In these two sections, we will present  some numerical results. 
In Appendix B, 
we will clarify the difference between  
two  boundary conditions 
 at a fixed potential difference and at fixed surface charges.   
In Appendix D, we will compare 
the electrostatics of  our discrete  
particle systems and   the continuum 
electrostatics.

\section{Charged particles} 

We consider   $N$ charged   particles  in a cell. 
Their positions and charges are 
 ${\bi r}_i$ and $q_i$ ($1\le i \le N)$. We assume 
the charge neutrality condition,   
\be 
\sum_i  q_i=0.
\en     
In terms of     the electrostatic energy  
$U= U({\bi r}_1, \cdots, {\bi r}_N)$, the  electrostatic 
force on    particle $i$ 
is given by 
\be 
{\bi F}_{i}^e= q_i {\bi E}_i = - \frac{\p}{\p {\bi r}_i}U ,
\en 
where   ${\bi E}_i$ is the electric field  on  particle $i$.  
There can also be neutral particles interacting with charged ones.

\subsection{Ewald method for charged particles  in the periodic boundary condition}
When  the bulk properties of charged particle  systems 
are calculated, the periodic boundary condition is  usually assumed 
in molecular dynamics simulation.  
In  a $L\times L\times L$ cell, the electrostatic energy 
  is written as \cite{Allen,Frenkel}  
\be
U_{\rm p}=\frac{1}{2} \sum_{\bi m}  {\sum_{i,j}}' 
\frac{q_iq_j}{|{\bi r}_{ij} + L{\bi m}|},
\en 
where $ {\bi r}_{ij} ={\bi r}_i-{\bi r}_j$ is the relative positional vector, 
${\bi m}=(m_x,m_y,m_z)$ is a 3D vector with three integer components, 
and the self terms ($j=i$ and ${\bi m}={(0,0,0)}$) are excluded 
in the summation $ {\sum_{i,j}}'$.  
Here,   ${\bi E}_i$ is obtained from Eq.(2.2)  with $U= U_{\rm p}$. 
In some papers\cite{Tyagi,Smith1},  discussions 
have been made on   
the convergence of the Ewald sum under Eq.(2.1).

In the Ewald method, the Coulomb potential 
$q_iq_j /r$ is divided into the short-range part 
$q_iq_j \psi_s(r)$ and the 
long-range part $q_iq_j \psi_\ell (r)$, where  
\be 
\psi_s(r)= {\rm erfc}(\gamma r)\frac{1}{r}, \quad 
\psi_\ell(r)= {\rm erf}(\gamma r)\frac{1}{r}. 
\en 
Here,  ${\rm erf}(u) = (2/\sqrt{\pi})\int_0^u du e^{-u^2}$ is the error 
function and ${\rm erfc}(u) =1-{\rm erf}(u)$ 
is the complementary error function.  The function $\psi_\ell(r)$ 
is the solution of $\nabla^2 \psi_\ell(r)= -4\pi \varphi_3(r)$, 
where  $\varphi_3(r)= 
\varphi(x)\varphi(y)\varphi(z)$ is a normalized 3D 
Gaussian distribution with 
\be 
\varphi(x)= (\gamma/\sqrt{\pi})  \exp( -\gamma^2 x^2).
\en 
The inverse $\gamma^{-1}$ is an adjustable  potential range. 
For $r\gg \gamma^{-1}$, we have  $\psi_\ell\cong r^{-1}$. 
For  $r\ll \gamma^{-1}$  it is finite  as  
\be 
\psi_\ell(r) =(2\gamma/\sqrt{\pi})(1- \gamma^2r^2/3\cdots ) .
\en 

 As is well-known \cite{Allen,Frenkel}, 
the Ewald form of  $U_{\rm p}$  reads  
\bea 
&&\hspace{-1cm} U_{\rm p}= 
\sum_{\bi m}  {\sum_{i,j}}' 
\frac{q_iq_j}{2} \psi_s ({|{\bi r}_{ij} + L{\bi m}|})
- \sum_i \frac{\gamma q_i^2}{\sqrt{\pi}}  \nonumber\\
&&\hspace{-1cm} +\frac{1}{L^3} {\sum_{{\bi \nu}\neq{\bi 0}}}\sum_{i,j} \frac{q_i q_j}{2}  
\Psi_\ell(k) e^{{\rm i}{\bi k}\cdot 
{\bi r}_{ij}} +\frac{2\pi}{3L^3} \bigg|\sum_i q_i {\bi r}_i\bigg|^2, 
\ena 
where the second term is equal to 
$-\sum_i q_i^2\psi_\ell(0)/2$ and   
 the last two terms 
arise from $\sum_{\bi m}  {\sum_{i,j}}  
{q_iq_j} \psi_\ell ({|{\bi r}_{ij} + L{\bi m}|})/2$ including the 
self part.  Use is made of  the Fourier transformation  
$\psi_\ell(r)= (2\pi)^{-3}
\int d{\bi k}\Psi_\ell(k) \exp[{{\rm i}{\bi k}\cdot{\bi r}}]$, where  
\be   
\Psi_\ell(k)=4\pi \exp(-k^2/4\gamma^2) /k^2.
\en  
In the third  term,  $\bi k$ is discretized as 
\be 
  {\bi k}= (2\pi/L) (\nu_x,\nu_y,\nu_z) , 
\en 
  and 
the term with  ${\bi \nu}=( \nu_x, \nu_y, \nu_z)= 
{(0,0, 0)}$ is excluded, where  $ \nu_x$, $\nu_y$, and $\nu_z$ 
are    integers ($0, \pm 1,\pm 2,\cdots$). 
This discretization stems from the summation over ${\bi m}$ in Eq.(2.3). 
The last  term in Eq.(2.7) arises from 
the  ${\bi k}$-integration  at   small 
$|{\bi k}|\ls L^{-1}$. 
However, it is   negligible  
without overall polarization or for $\sum_i q_i{\bi r}_i\cong {\bi 0}$. 
See Appendix A  for the  derivation of 
the last two terms.

\subsection{Ewald method  for charged particles between metallic plates under applied electric field}

As in Fig.1, we consider 
    a  $L\times L\times H$ parallel plate geometry,  
where the plates at $z=0$ and $H$ are both metallic. 
They are assumed to be smooth and  structureless. 
 We assume the periodic boundary condition 
along the $x$ and $y$ axes. We   adopt  the  fixed-potential boundary 
condition. See Appendix B for discussions of 
another typical  boundary condition with  fixed surface charges.

We treat  an infinite number of  image 
charges. Let   
 a particle $i$  at ${\bi r}_i=(x_i, y_i, z_i) $  
with charge $q_i$ approach the surface  at $z=0$. 
For not strong applied electric field, 
it is acted  by a growing  attractive force  produced 
 by the  closest image   with the opposite charge  $-q_i $ 
 at \cite{Landau}, 
\be 
{\bar{\bi r}}_i=(x_i, y_i, -z_i).
\en 
This  force is written as    
  $-\p v_{\rm I}/\p z_i = -q_i^2/(2z_i)^2$  for small $z_i$,  
so   the  potential due to this image   is of the form, 
\be    
  v_{\rm I}(z_i) = -q_i^2/4z_i.
\en   

\begin{figure}
\includegraphics[width=0.75\linewidth]{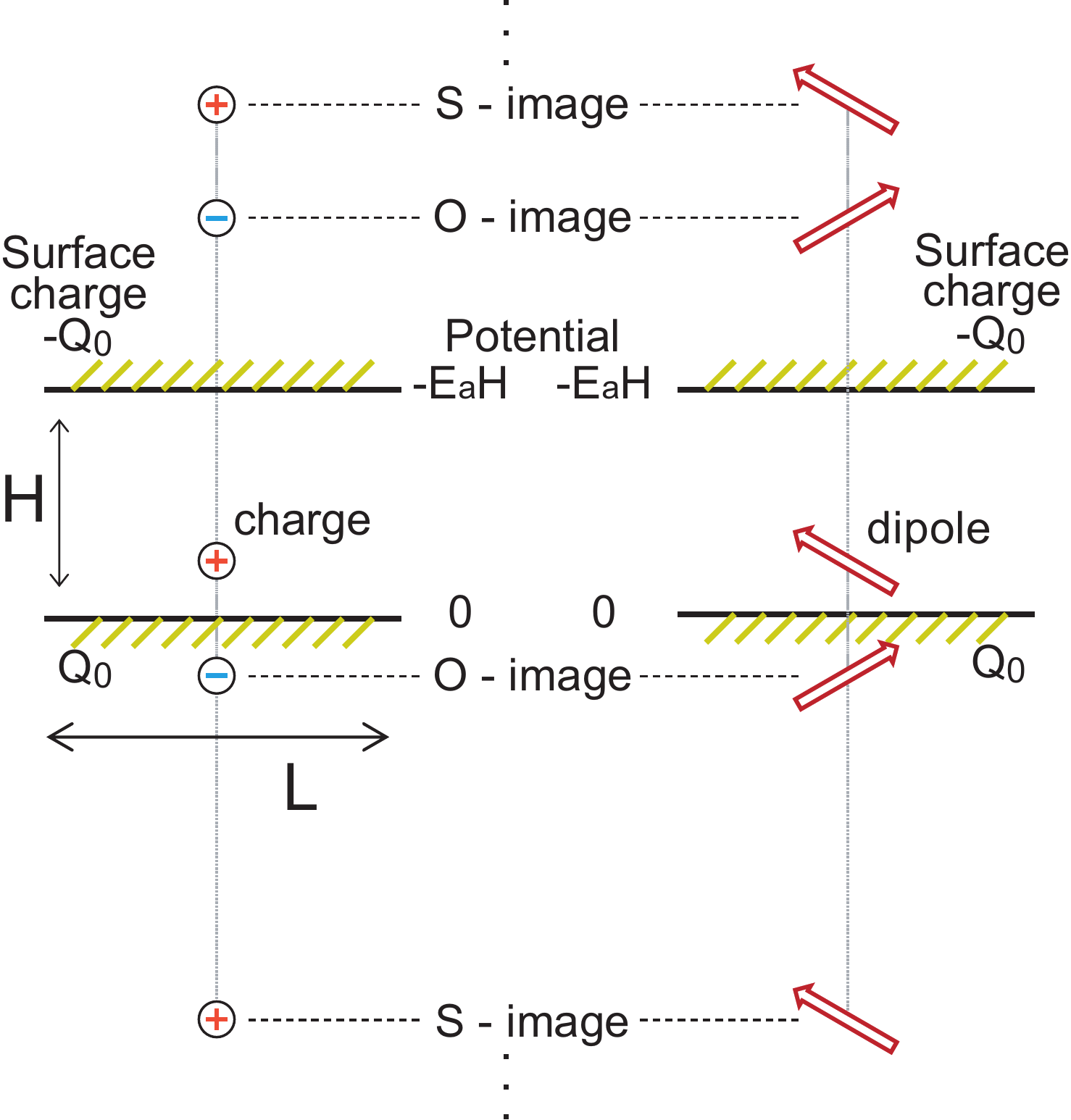}
\caption{ (Color online) 
System of a capacitor with parallel 
metallic plates containing  
charges (left)  and  dipoles (right). 
For each charge or dipole  at $(x_j, y_j, z_j)$, 
there appear O-images  
at $(x_j, y_j, -z_j-2Hn)$ ($n=0, \pm 1, \cdots)$ 
and S-images at  $(x_j, y_j, z_j-2Hn)$ ($n= \pm 1, \cdots)$.  
 The  electrostatic potential $\Phi$ is  $0$ 
at the bottom and is    $-\Delta\Phi= -E_aH$ at the top. 
The  surface charge is $Q_0$ at the bottom and $-Q_0$ at the top. 
As typical  experiments, 
we may fix either of $\Delta\Phi$  or $Q_0$.}
\end{figure}

\subsubsection{Electrostatic potential and image charges}

We   write the electrostatic potential 
away from the particle positions $({\bi r} \neq {\bi r}_j$) 
as 
\be 
\Phi({\bi r})= \phi({\bi r})-E_a z.   
\en   
The electric field away from the particle positions 
is ${\bi E}= E_a{\bi e}_z-\nabla\phi$, where ${\bi e}_z$ is the unit vector 
along the $z$ axis.  
We assume that  the charged particles are 
repelled  from the  walls at short  distances 
and that  no  ionization occurs on the walls. 
Then, the excess potential 
$\phi$ in Eq.(2.12)  satisfies the boundary condition,  
\be 
\phi (x,y,0)=\phi(x,y,H)=0.  
\en 
From the bottom to the top, the potential difference is  
\be 
\Delta \Phi= \Phi(x,y,0)- \Phi(x,y,H)=HE_a.
\en

In the region $0<z<H$, $\phi({\bi r})$ is the solution of 
the Poisson equation  under the boundary condition (2.13), 
\be 
-\nabla^2\phi= 
4\pi \sum_j q_j \delta({\bi r}- {\bi r}_j).  
\en 
To calculate $ \phi({\bi r})$, 
we consider its  2D  Fourier expansion,    
\be 
\phi=
 \frac{4\pi}{L^2}\sum_{{\bi \nu}_\perp}\sum_j q_j 
G_k (z,z_j)   e^{{\rm i}\cdot(
k_x (x-x_j)+k_y(y-y_j))} , 
\en 
where ${\bi \nu}_\perp= (\nu_x, \nu_y)$   and 
$(k_x,k_y)= (2\pi /L)(\nu_x,\nu_y)$  
 with $\nu_x$ and $\nu_y$ being integers.  
  From Eq.(2.15) 
the Green function  $G_k (z,z')$ satisfies  
\be 
(k^2- \p^2/\p z^2) G_k(z,z') = \delta (z-z'), 
\en 
where  $k= (k_x^2+k_y^2)^{1/2}$. Here,  
 $G_k(0,z')=G_k(H,z')=0$   from Eq.(2.13), so we 
 solve Eq.(2.17) as \cite{OnukiNATO,Perram}    
\bea 
G_k(z,z')&=& \frac{1}{2k} e^{-k|z-z'|}-\frac{1}{2k\sinh (kH)} 
 \nonumber\\ 
\vspace{0.5mm}
&&\hspace{-2.7cm} \times 
 \bigg [ \sinh (kz)e^{-k(H-z')} +   \sinh (kH-kz)e^{-kz'}  
\bigg ],
\ena
which has  the symmetry $G_k(z,z')=G_k(z',z)$.
The first term in Eq.(2.18) arises from the 2D Fourier transformation of 
 the direct Coulombic  interaction, while the second term 
is induced by the surface charges on the metallic surfaces. 
Note that   the term   
with $k_x=k_y=0$ is included in Eq.(2.16), 
which give rise to a term independent of $x$ and  $y$.  
From Eq.(2.18) 
 the long wavelength limit 
$G_0(z,z')=\lim_{k\rightarrow 0} G_k(z,z')$ becomes 
\be 
G_0(z,z')= ( z+z'-|z-z'|)/2 -  zz'/H   . 
\en

To find  image charges, we further use 
  the expansion $1/\sinh(kH) 
=2\sum_{n\ge 0} e^{-kH(2n+1) }$ in Eq.(2.18)  
to obtain    
\be 
\hspace{-1mm} 
G_k(z,z')=\frac{1}{2k} 
\sum_{n}\bigg 
[e^{-k|z-z'+2Hn|}-e^{-k|z+z'+2Hn|}\bigg], 
\en 
where $n =0,\pm 1, \pm 2,\cdots$. Since  
 $ 2\pi \exp({-k|z|})/k= \int dxdy 
\exp(ik_xx+ik_yy)/ r$,  
substitution of the above expansion into Eq.(2.16) 
 yields  $\phi({\bi r})$  in the 
following  superposition of  Coulomb potentials, 
\be 
\phi =  \sum_{{\bi m}} {\sum_j}
\frac{q_j}{|{\bi r}-{\bi r}_{j} + 
{\bi h}|} -    \sum_{\bi m} {\sum_j}
\frac{q_j}{|{\bi r}-{\bar{\bi r}}_{j}  
 + {\bi h}|} ,
\en   
where 
 ${\bi m}= (m_x,m_y,m_z)$ with integer components  and  
\be 
{\bi h}= (Lm_x,Lm_y,2Hm_z).
\en 
For  each charge $q_j$ at ${\bi r}_j=(x_j,y_j,z_j)$ in the cell, 
we find  images with  
the opposite charge  $ -q_j $ at 
$(x_j, y_j, -z_j-2Hn)$ ($n=0, \pm 1, \cdots)$, giving rise to the second term 
in Eq.(2.21). We call them  O-image charges. 
We also find those    with the same  charge  $q_j$ 
at   $(x_j, y_j, z_j-2Hn)$ ($n= \pm 1, \cdots)$ 
in the first term in Eq.(2.21), which are called S-image charges. 
See Fig.1 for these images.

The electrostatic energy  
$U_{\rm m}$ is now given by  \cite{Hautman,Perram}  
\bea 
U_{\rm m} &=& 
 \frac{1}{2} \sum_{\bi m} {\sum_{ij}}'
\frac{q_iq_j}{|{\bi r}_{ij} + 
{\bi h}|}  
-   \frac{1}{2} \sum_{\bi m} {\sum_{ij}}
\frac{q_iq_j}{|{\bar{\bi r}}_{ij}  
 + {\bi h}|}  \nonumber\\
&&-E_a \sum_i q_i z_i,
\ena  
where ${\bar{\bi r}}_{ij}={\bi r}_i- 
{\bar{\bi r}}_j= (x_i-x_j,y_i-y_j, z_i+z_j)$. 
In  $\sum_{ij}'$, 
we exclude the self term with $j=i$ 
for  ${\bi m}=(0,0, 0)$. 
The second term includes the 
direct image potential (2.11).  
 For infinitesimal changes 
${\bi r}_i \to {\bi r}_i+ d{\bi r}_i$ and $E_a \to E_0+dE_a$, 
 the differential form of $U_{\rm m}$ in Eq.(2.23)  is  given by  
\be
d U_{\rm m} =  -\sum_i q_i {\bi E}_i \cdot d{\bi r}_i 
- \sum_i q_i z_i   dE_a .
\en

\subsubsection{Ewald representation}

To obtain the Ewald representation,  
we divide  the Coulomb potentials  
 into the short-range and long-range parts as in  Eq.(2.4). 
We then find     
\bea 
&&\hspace{-6mm}  
U_{{\rm m}} 
  =\sum_{\bi m} \bigg[  {\sum_{i,j}}'
\frac{q_iq_j}{2} \psi_s({|{\bi r}_{ij} + 
{\bi h}|}) -  {\sum_{i,j}}
\frac{q_iq_j}{2} \psi_s({|{\bar{\bi r}}_{ij} 
 + {\bi h}|})\bigg]
\nonumber\\
&&-  \sum_i \frac{\gamma q_i^2}{\sqrt{\pi}}  
+ U_{\rm m}^\ell -E_a \sum_i q_i z_i,
\ena 
where the first term is the short-range part and $U_{\rm m}^\ell$ is the 
long-range part (including the self terms) given by 
\bea
&&  U_{\rm m}^\ell= 
\sum_{\bi m}  {\sum_{i,j}} 
\frac{q_iq_j}{2} \bigg[ \psi_\ell({|{\bi r}_{ij} + 
{\bi h}|}) -   \psi_\ell({|{\bar{\bi r}}_{ij} 
 + {\bi h}|})\bigg]
\nonumber\\
&&= \frac{1}{HL^2} 
{\sum_{{\bi \nu}_\perp\neq {\bi 0}}}
\sum_{i,j} \frac{q_i q_j}{4}    
\Psi_\ell(k) [ e^{{\rm i}{\bi k}\cdot 
{\bi r}_{ij}}-  e^{{\rm i}{\bi k}\cdot 
{\bar{\bi r}}_{ij}}] \nonumber\\
&&\hspace{4mm} +  \frac{2\pi}{L^2}\sum_{i,j} q_iq_j {K}_0(z_i,z_j) .
\ena 
Here,  the wave vector $\bi k$ is discretized as 
\be 
{\bi k}= \pi(2 \nu_x/L,2 \nu_y/L,  \nu_z/H),   
\en 
where 
$\nu_x, \nu_y, $ and $\nu_z$ are  integers. 
The  summation in Eq.(2.26) is 
over ${\bi\nu}=(\nu_x,\nu_y,\nu_z)$ with 
${\bi \nu}_\perp=(\nu_x,\nu_y) \neq (0,0)$. 
The last term in Eq.(2.26) arises  from 
the {\it one-dimensional} contributions with  
   $\nu_x=\nu_y=0$ and $\nu_z\neq 0$, where the contribution from 
${\bi\nu}=(0,0,0)$ vanishes in the present case
\cite{Hautman}. 
In  Appendix C,  we will derive 
\bea 
&&\hspace{-5mm}K_0(z,z')= \int_0^H du 
G_0(z,u)[{\hat\varphi}(u-z')-{\hat\varphi}(u+z')] \nonumber\\ 
&&\hspace{-9mm}
= \frac{2H}{\pi^2} \sum_{n \ge  1} \frac{\exp[- (\pi n/2\gamma H)^2]}{n^2}  
\sin(\frac{\pi nz}{H})\sin(\frac{\pi nz'}{H}), 
\ena 
where   $G_0(z,u)$ is given  by   Eq.(2.19) and 
 $\hat{\varphi}(z)$ is a periodic function with period $2H$. 
Use of  $\varphi(z)$ in  Eq.(2.5) gives  
\be 
\hat{\varphi}(z)= \sum_{n=0,\pm 1, \cdots}\varphi(z-2Hn), 
\en 
so   $\hat{\varphi}(z)= \hat{\varphi}(-z)$ and 
  $\int_0^H d{z}\hat{\varphi}(z)=1/2$. 
From the first line  of Eq.(2.26), 
  $U_{\rm m}^\ell$  vanishes as 
$z_i$ (or $z_j$) tends to $0$ or  $H$, 
while   the first term 
 in Eq.(2.25) diverges in this  
limit since it contains $v_{\rm I}(z_i)$  in Eq.(2.11).

\subsubsection{Surface charges }

In the parallel plate geometry, real charges are those within the cell 
and the excess electrons 
on the metal  surfaces  at $z=0$ and $H$. 
The image changes are introduced  as a mathematical convenience. 
The surface charge densities   are 
 given by $\sigma_0(x,y)={E_z(x,y,0)}/4\pi$ at $z=0$ 
and   $\sigma_H(x,y)=-{E_z(x,y,H)}/4\pi$ at $z=H$, where 
$E_z(x,y,z)= -\p \Phi/\p z$.   
From Eq.(2.18) we obtain their  2D Fourier expansions,  
\bea 
&&\hspace{-6mm}\sigma_0=\frac{{E_a}}{4\pi} - 
 \frac{1}{L^2}\sum_{{\bi \nu}_\perp}\sum_j q_j 
\frac{\sinh(kH-kz_j)}{\sinh(kH)} 
   e^{{\rm i}{\bi k}\cdot({\bi r}- 
{\bi r}_j)} ,\nonumber\\
&&\hspace{-6mm}\sigma_H=  - \frac{{E_a}}{4\pi} - 
 \frac{1}{L^2}\sum_{{\bi \nu}_\perp}\sum_j q_j 
\frac{\sinh(kz_j)}{\sinh(kH)}  e^{{\rm i}{\bi k}\cdot({\bi r}- 
{\bi r}_j)}.  
\ena 
Here,  ${\bi \nu}_\perp=(\nu_x,\nu_y)$,    
${\bi k}=(k_x,k_y,0) = 
 (2\pi/L)(\nu_x,\nu_y,0)$, and  
  $\exp[{\rm i}{\bi k}\cdot({\bi r}- 
{\bi r}_j)]=
\exp[{\rm i}(k_x (x-x_j)+{\rm i} (k_y(y-y_j)]
$.

The total electric charges  on the bottom and top surfaces 
$Q_0$ and $Q_H$ are the surface integral  
of $\sigma_0(x,y)$   and  that of 
 $ \sigma_H(x,y) $, respectively. 
We  pick  up the terms with ${\bi \nu}_\perp=(0,0)$ 
in Eq.(2.30) to obtain     
\be 
Q_0= -Q_H= \frac{1}{4\pi} 
L^2E_a +   \frac{1}{H}\sum_j q_jz_j . 
\en 

\subsubsection{Local electric field }

From Eqs.(2.25) and (2.26) we may calculate 
 the local electric field ${\bi E}_i$ on particle $i$ 
using Eq.(2.2). The contribution from the first term in Eq.(2.25) 
is written as 
${\bi E}_{i}^{\rm s}$, that from the first term in Eq.(2.26) 
 as ${\bi E}_i^{\ell }$, and 
that from the  remaining   terms as 
$E_i^{0 }{\bi e}_z$.   Then, 
\be  
{\bi E}_i = {\bi E}_{i}^{\rm s}+ {\bi E}_{i}^{\ell }+ 
{E}_{i}^{0}{\bi e}_z.
\en 
Here,   $E_i^{0 }$ is determined by the $z$ coordinates, 
 $z_1, z_2, \cdots$, and use of 
 Eqs.(2.28) and (C3) gives   
\bea
&&\hspace{-4mm}
E_i^0=\frac{4\pi}{L^2}\bigg[Q_0-  \sum_{j}q_j 
 \int_{-z_j}^{z_j}\hspace{-1mm} du 
{\hat\varphi}(u-z_i) \bigg]   \nonumber\\
&&\hspace{-5mm} =  E_a- \frac{8}{L^2}
\sum_{n \ge 1} e^{- ({\pi n}/{2\gamma H})^2}  \frac{J_n}{n}
 \cos(\frac{\pi nz_i}{H}),
\ena  
where  $J_n$ depends on $z_1, z_2, \cdots$ as 
\be 
J_n= \sum_j {q_j}   \sin(\frac{\pi nz_j}{H}).
\en 
The first line of Eq.(2.33) tends to 
a well-defined  limit 
 in the continuum theory 
in Appendix D. 
Note that the last term of Eq.(2.26) 
is expressed as $(4H/\pi L^2) \sum_{n\ge 1} 
e^{- ({\pi n}/{2\gamma H})^2} J_n^2$ in terms of $J_n$.



\subsection{Numerical example of charged particles between metallic plates under applied electric field}

\subsubsection{Model and method}
This subsection presents   results of  molecular dynamics simulation of 
  two-component charged particles  
between  metallic plates at the fixed-potential condition.  
The numbers  of the two  species are  $N_1=N_2=500$ and 
the charges    are  $q$ and $-q$. 
 All the particles have a common mass $m$ and a common diameter $\sigma$. 
The average density is $n_0= (N_1+N_2) / HL^2 =0.57 \sigma^{-3}$ 
and the cell dimensions are $H=L=12\sigma$. 

The  total potential energy consists of three parts as 
\be 
U_{\rm tot}= U_{\rm m} 
+ \sum_{i>j}v_s(r_{ij})  +\sum_i v_{\rm w}(z_i).
\en 
First,   $U_{\rm m}$ is given by  Eq.(2.25), 
where we set  $\gamma=0.5 $ 
and sum over $\bi k$  
in the region  $k\le k_c=18 \pi/L$.  
From  the second  line of  Eq.(2.28), 
we calculated the one-dimensional electric field  $E_i^0$ 
retaining the terms up to $n=18$, for which there is 
almost no error since   
 $\exp[-(\pi n/2\gamma H)^2] \sim 10^{-10}$ at $n=18$. 
We also truncated 
the short-range part of the 
electrostatic interaction in  Eq.(2.25)  at   $4\sigma$. 
Second,  $v_s(r)$ is the  soft-core pair potential,  
\be
v_{s}(r) =4\epsilon ({\sigma}/{r})^{12} -C_0,
\en 
where $\epsilon$ is the characteristic 
 interaction energy. 
This potential is cut off at 
 $r= r_{\rm cut}^s=4\sigma$ 
and $C_0$ ensures  the continuity of $v_s(r)$  
at this  cut-off. Third, $v_w(z)$ is the 
 repulsive potential from the walls written as 
\be 
v_w(z)=w\exp(-z/\xi)+w\exp(-(H-z)/\xi),
\en 
with  $\xi=0.01\sigma$ and $w=e^{40}\epsilon$. 
The minimum of $z_i$ and $H-z_i$ was 
then about $0.4\sigma$ in our examples.

In this paper, the charge size is set equal to
\be 
q= |q_i|= 5 ({\epsilon\sigma})^{1/2}. 
\en    
Hereafter, units of length, density, electrostatic 
potential, electric field, 
and electric charge   are $\sigma$, $\sigma^{-3}$, $(\epsilon/\sigma)^{1/2}$, 
 $(\epsilon/\sigma^3)^{1/2}$, 
and $(\epsilon\sigma)^{1/2}$, respectively.
The temperature $T$ is in units of  $\epsilon/k_B$. 
Due to  the choice (2.38), the  typical magnitude 
of the Coulomb interaction among the particles 
is   $q^2n_0^{1/3} \sim 20$ per particle, while   
that  of the image interaction in Eq.(2.11) is   
 $10$ at  $z_1\sim 0.5$.

The  particles obey the equations  of motions,  
\be 
m   \frac{d^2}{dt^2}{{\bi r}}_i =-\frac{\p}{\p{\bi r}_i} 
 U_{\rm tot}.  
\en 
We measure  time in units of  
\be 
\tau_0= (m/\epsilon)^{1/2}\sigma.
\en 
We integrated Eq.(2.39)  using  
the leap-frog method with the time step width being 
$0.002$.  With   a Nos\'e-Hoover thermostat  attached, 
we started  with a high temperature liquid state, 
lowered   $T$  to a final value,   
and  waited for a time interval of 
$10^4$. Next, we removed the thermostat, waited 
 for another  time interval of 
$10^4$, and took  data.  Thus, our simulation results  
 were those  in the NVE ensemble, where 
the average kinetic energy was kept at  
 $3k_BT/2$ per particle.
Hereafter, $T$ has this meaning.
In this paper, we treat equilibrium states
with homogeneous $T$.

We  also carried out  simulations 
for  ($\gamma,r_{\rm cut}^s)=(0.3, 6)$ 
and $(1,6)$ for the situations in Fig.2.
The results were 
 essentially the same as 
those for the present choice $(0.5, 4)$, while the long-range contributions 
(including $E_i^0$) become
 smoother  with decreasing $\gamma$. 
 See  discussions  
on the choice of $\gamma$  by 
 Deserno and Holm \cite{mesh}.

\begin{figure}
\includegraphics[width=0.96\linewidth]{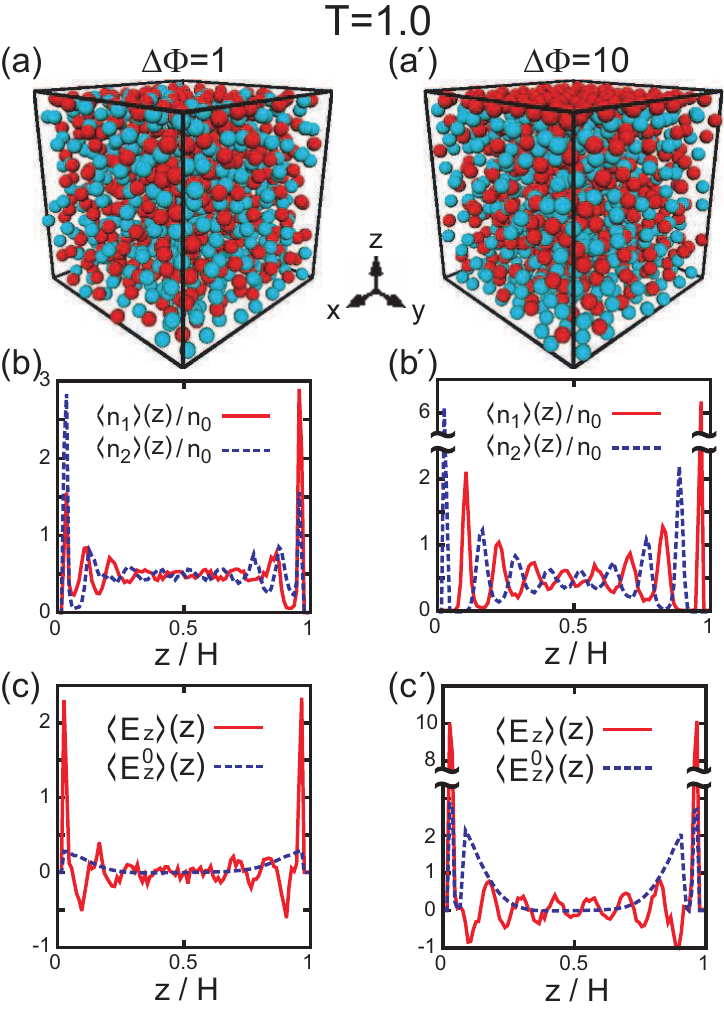}
\caption{(Color online)  Simulation results of 1000 charged particles 
in liquid at $T=1$. They have a common diameter $\sigma$   
in a cubic cell with length $12$. 
Snapshots of  
cations (in red) and anions (in blue)  
for  $\Delta \Phi=1$  in (a)  and $10$ in (a$'$). 
Lateral density averages $\av{n_1}(z)$  and 
$\av{n_2}(z)$  divided by the space average $n_0$ 
are in (b) and (b$'$). 
Lateral averages  $\av{E_{z}}(z)$ (red bold line)  
 and  $\av{E_{z}^0}(z)$ (blue dotted line)  are in (c) and (c$'$). 
Particles are attracted to the walls and screening occurs. 
}
\end{figure} 

\begin{figure}
\includegraphics[width=0.96\linewidth]{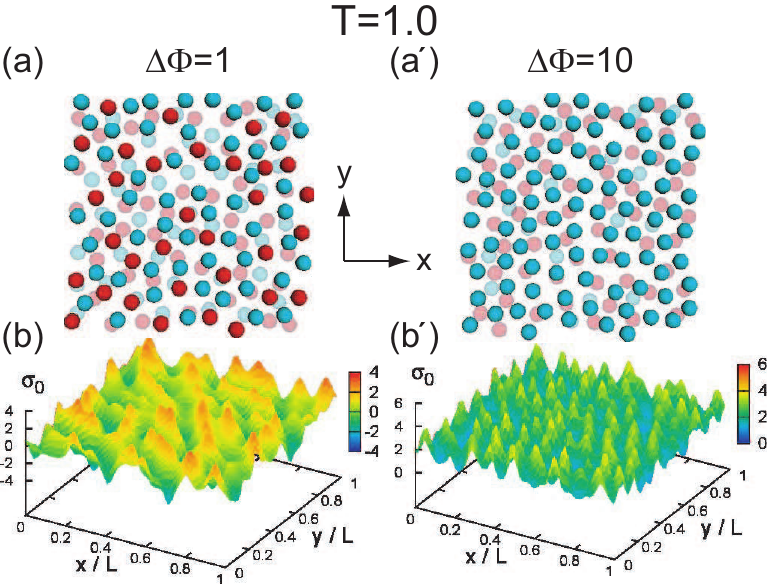}
\caption{(Color online) Top: Particle configurations  
in the first layer $0<z<1$ (in red or blue) 
 and in the second layer $1<z<2$  
(in lighter colors) 
 for $\Delta \Phi=1$ (left) and  $10$ (right).  
Bottom: Surface charge density  
$\sigma_0(x,y)$ 
at $z=0$ in Eq.(2.30) for  $\Delta \Phi=1$ and $10$  
in units of $(\epsilon/\sigma^3)^{1/2}$. 
Data are common to those  in (a) and (a$'$) of  Fig.2. 
}
\end{figure} 

\subsubsection{Charged particles at $T=1$: Screening effect} 

In  Figs.2 and 3,  we show simulation results  in liquid  at 
$T=1$ for $\Delta\Phi=1$ and $10$. In Fig.2(a)-(a$'$), snapshots 
of the particles are given. 
In Fig.2(b)-(b$'$),   cross-sectional densities  
$ \av{n_1} (z)$ and  $\av{n_2} (z)$ 
 are displayed, which  show  accumulation  of 
the negative (positive) charges near  the wall at $z=0$ ($z=H$).  
They are defined as follows. 
Let $\Delta N_\alpha(z)= 
\sum_{j\in \alpha}{\theta(z_j-z)\theta(z+\Delta z-z_j)}$  
be the particle numbers in  layers   $[z,z+\Delta z]$ 
($z= n\Delta z$ with $n=0,1,\cdots)$   
for the two species $\alpha=1,2$. The $\theta(u)$ is the step function being 1 for $u>0$ and 0 for $u\le 0$.  In this subsection, 
 we set $\Delta z=H/100=0.12$, which is much smaller than 
the particle size. 
The laterally averaged densities  for the two species are given by 
\be 
 \av{n_\alpha} (z)=\av{\Delta N_\alpha(z,t)}/L^2\Delta z. 
\en 
Hereafter,  $\av{\cdots}$ represents the average over a time interval 
with  width $10^3$.

In Fig.2(c)-(c$'$), 
we display  the   laterally averaged (local) electric field 
along the $z$ axis  calculated from   
\be 
\av{E_z}(z)= 
\sum_j\av{ E_{zj} \frac{\theta(z_j-z)\theta(z+\Delta z-z_j)}{\Delta N(z)}},
\en 
where  
$q_j E_{zj}=- \p U_{\rm m}/\p z_j$ and 
$\Delta N(z)=\Delta N_1(z)+\Delta N_2(z)$. 
  We can see that 
 $\av{E_z}(z)$ exhibits sharp peaks  near the walls but 
tends to zero in the interior.  This 
screening is achieved only by 
one or two  layers of the accumulated charges. 
In the same manner, replacing   $E_{zj}$  in Eq.(2.42) by $E_{zj}^s$, 
$E_{zj}^\ell$, and $E_{j}^0$  in Eq.(2.33),  
 we may  define the  lateral averages   
$\av{E^s_z}(z)$, $\av{E_z^\ell }(z)$, 
and $\av{E^0}(z)$,  respectively. 
For  the examples in this subsection, we find  
$\av{E_z^\ell}(z)\cong 0$  (less than 0.01) so that 
\be 
\av{E_z}(z)\cong \av{E_z^s }(z) + \av{E^0_z}(z). 
\en 
In  Fig.2(c)-(c$'$),  $\av{E^0}(z)$ 
 more smoothly varies than $ \av{E_z}(z)$  
and tends to zero  far from the walls,  while  
$\av{E_z^s}(z)$ exhibits sharp peaks 
near the walls and oscillates 
around zero far from the walls.

\begin{figure}[t]
\includegraphics[width=0.97\linewidth]{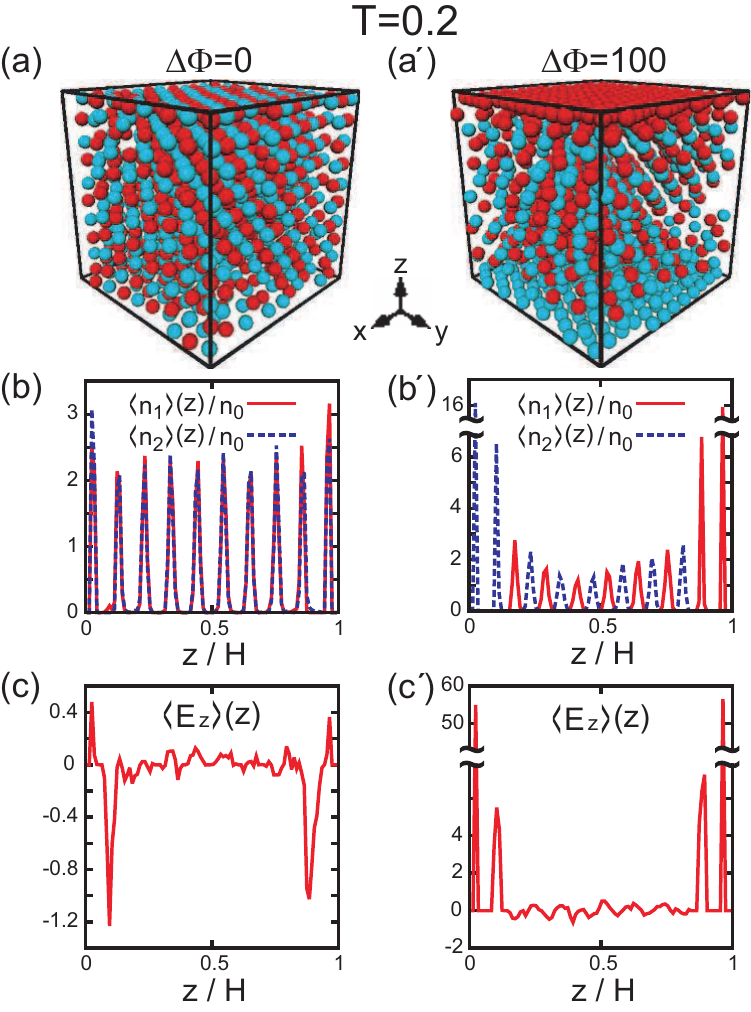}
\caption{(Color online) 
 Simulation results of charged particles in crystal  at $T=0.2$. 
Cations (in red) and anions (in blue) 
for $\Delta \Phi=1$ in (a) and $100$ in (a$'$). Normalized  
lateral density averages $\av{n_1}(z)/n_0$  and 
$\av{n_2}(z)/n_0$ are in (b) and (b$'$). 
Lateral average $\av{E_{z}}(z)$ are in (c) and (c$'$). For $\Delta\Phi=0$,  
both  cations and anions are somewhat  denser 
in the first layer with larger  $\av{E_{z}}$ with a negative  
 $\av{E_{z}}$  in the second layer. 
For $\Delta\Phi=100$, anions are  closely packed in the first layer, 
while their lateral density gradually 
decreases from the second layer.   
These ordered configurations are nearly stationary. 
}
\end{figure} 
\begin{figure}
\includegraphics[width=0.96\linewidth]{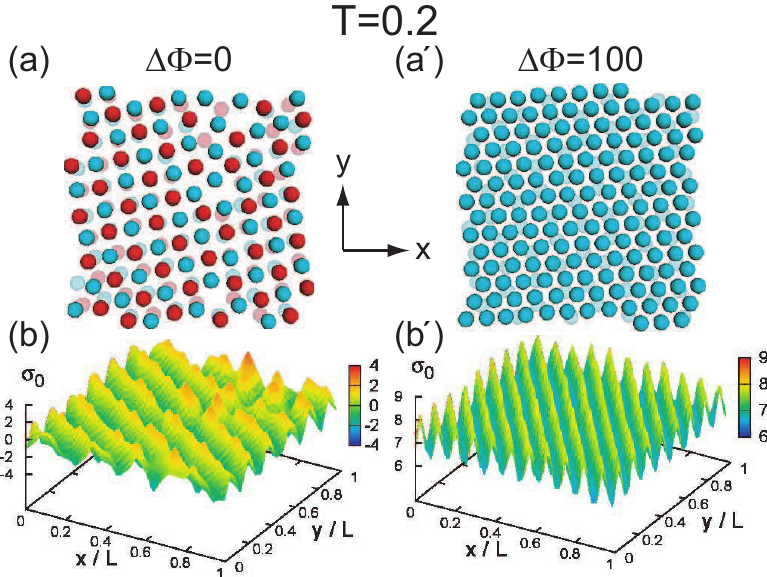}
\caption{(Color online) 
Particle configuration 
in the first layer $0<z<1$ (in red or blue) 
 and in the second layer $1<z<2$  (in lighter colors) 
 for $\Delta \Phi=0$ in (a)  and  $100$ in (a$'$).
 Surface charge density  
$\sigma_0(x,y)$ at $z=0$ for  $\Delta \Phi=0$ in (b) and $100$ in (b$'$) 
in units of $(\epsilon/\sigma^3)^{1/2}$. 
 Data are common to those  in (a) and (a$'$) of  Fig.4. 
For $\Delta\Phi=0$,  
a square lattice (with defects) appears from the 
first layer. 
For $\Delta\Phi=100$, 
anions form a hexagonal lattice  in the first layer, but  
are in a transition state in the second layer. 
 }
\end{figure}

In Fig.3, we  show snapshots of the particles  
in the two layers $0<z<1$ and $1<z<2$. 
For $\Delta\Phi=1$, cations and anions 
are both   attracted  to the wall 
due to the image interaction. For $\Delta\Phi=10$, 
only anions  are attached in the first layer, but 
cations are richer  in the second layer.  

We divide   Eq.(2.31) by $qL^2$  to obtain  
\be 
 Q_0/qL^2= E_a/4\pi q 
+ \sum_j q_jz_j/qHL^2,
\en 
where the second term is    the  areal 
 density of excess cations  
at the top (with $z_j \cong H$) 
and is larger than 
 the first term for strong screening 
 \cite{commentS}. The data  in Figs.2(a)-(a$'$) give 
 $(E_a/4\pi q,   \sum_jq_jz_j /qHL^2)= 
(0.0013, 0.106)$  for $\Delta\Phi =1$ 
and $(0.0013, 0.426)$  for $\Delta\Phi  =10$.

 \subsubsection{Charged particles at $T=0.2$: Ionic crystals} 
 
In  Figs.4 and 5,  we show simulation results  in crystal   at 
$T=0.2$.  For $\Delta\Phi=0$  
the cations and anions 
 form a square lattice, as is well known for salts such as NaCl, 
 from  the first layer.  
However,  for     $\Delta\Phi=100$, we can see 
a hexagonal structure   in the first layer  $0<z<1$ and 
 a  square lattice  for $z>2$, 
where the  structure changes over 
in the second layer $1<z<2$.  
In this case, the first layer is composed of anions 
only and the second layer is anion-rich, as can also be seen in Fig.4(b$'$). 
In the lower plates of Fig.5, 
the surface charge 
density $\sigma_0(x,y)$ varies  nearly in one direction 
as a result of  the complex  
 charge distributions  in the first few layers.

In addition, we examine  Eq.(2.44). From the data  in Figs.4(a)-(a$'$)  
we have  $(E_a/4\pi q,   \sum_jq_jz_j /qHL^2)= 
(0, 0.024)$ for $ \Delta\Phi=1$  
and $(0.132, 1.31)$ for $ \Delta\Phi=100$.  In the  second example,    
the first layers next to  the walls are  nearly closely packed 
with cations or anions,  as can be seen in  Fig.5(a$'$).

\section{Polar particles} 
\setcounter{equation}{0} 
In this section, 
 we  consider    $N$ polar particles at positions ${\bi r}_j$ 
and with dipole moments ${\bi \mu}_j$. 
The derivatives of the electrostatic energy $U$ 
give the electrostatic force ${\bi F}_i^e$ and 
 the electric field ${\bi E}_i$   on 
dipole  $i$ as 
\be
{\bi F}_i^e= - \frac{\p}{\p {\bi r}_i}U, 
\quad {\bi E}_i= -\frac{\p}{\p {\bi \mu}_i}U,
\en  
which  are used in the equations of motions.  
There can also be neutral and charged 
particles.

\subsection{Ewald method for dipoles in  the periodic boundary condition}

We first consider  the electrostatic energy 
in the periodic boundary condition 
\cite{Allen,Frenkel,Leeuwreview}, which is 
written as  $ U_{\rm p}^{\rm d}$. Here, there is no 
 applied electric field. In  Eq.(2.3) we replace 
 $q_i$  by ${\bi \mu}_i\cdot\nabla_j$, where 
$\nabla_i=\p/\p {\bi r}_i$. Then, we obtain 
\be 
U_{\rm p}^{\rm d}= \frac{1}{2} 
  \sum_{\bi m} {\sum_{i,j}}'({\bi \mu}_i \cdot\nabla_i) 
({\bi \mu}_j \cdot\nabla_j) 
\frac{1}{|{\bi r}_{ij} + L{\bi m}|}. 
\en 
From Eq.(2.7) the Ewald representation is given by
\bea 
&& U_{\rm p}^{\rm d}= \frac{1}{2}
\sum_{\bi m}  {\sum_{i,j}}' 
{\bi \mu}_i \cdot  {\aw}_s ({|{\bi r}_{ij} + L{\bi n}|}) \cdot{\bi\mu}_j 
\nonumber\\
&&\hspace{-1mm} + \frac{1}{2L^3}   {\sum_{{\bi k}}}'\sum_{i,j} 
 ({\bi k}\cdot{\bi \mu}_i)  ({\bi k}\cdot{\bi \mu}_j)\Psi_\ell(k) 
 e^{{\rm i}{\bi k}\cdot 
{\bi r}_{ij}}   
\nonumber\\
&& - \sum_i \frac{2\gamma^3 }{3\sqrt{\pi}} |{\bi \mu_i}|^2
+\frac{2\pi}{3L^3}\bigg|\sum_{j}{\bi \mu}_j\bigg|^2.  
\ena 
In the first term, we introduce   the tensor  ${\aw}_s({\bi r})$ by   
\be
\aw_s({\bi r}) = -\nabla\nabla \psi_s(r) 
=B(r) \aI -  C(r) \frac{1}{r^2} {\bi r}{\bi r} ,
\en  
where  $\aI$ is  the unit tensor and  
\bea
&&\hspace{-8mm}B(r)=\psi_s(r)/r^2 + 2\pi \varphi_3(r)/\gamma^2r^2,
 \nonumber\\
&&\hspace{-8mm}C(r)
= 3B(r)+ 4\pi \varphi_3(r),
\ena
with  
$\varphi_3(r)= \varphi(x) \varphi(y) \varphi(z) $ (see Eq.(2.5)).
  The third term in Eq.(3.3) is 
the counterpart of the second term in Eq.(2.7) 
following from the small-$r$ expansion (2.6).

\subsection{Ewald method for dipoles between metallic  plates under applied electric field}

We consider $N$ dipoles between metallic plates 
in   applied electric field $E_a$. 
The electrostatic potential $\Phi$ is written as Eq.(2.12). 
For  each dipole 
 ${\bi\mu}_i= (\mu_{xi}, \mu_{yi},\mu_{zi})$ at ${\bi r}_i=(x_j,y_j,z_j)$, 
there arise  S-image dipoles   
 at $ (x_i,y_i, z_i-2Hn)$   ($n \neq 0$) 
with the same moment  ${\bi\mu}_i$
 and  O-image dipoles   at 
 $ (x_i,y_i, -z_i-2Hn)$  with the moment,
\be 
\bar{\bi\mu}_i= 
(-\mu_{xi}, -\mu_{yi},\mu_{zi}).
\en 
See Fig.1 for these images. 
If a dipole at  ${\bi r}_i$ approaches the bottom wall, 
its image at ${\bar{\bi r}}_i$ yields  the 
dipolar  electric field 
$- \p v_{\rm I}^{\rm d}/\p{\bi \mu}_i = -(2z_i)^{-3} 
(\bar{{\bi \mu}}_i- 3\bar{\mu_{zi}}{\bi e}_z)$ at ${\bi r}_i$. 
Thus the  interaction energy from this  closest image is 
written as      
\bea 
&&\hspace{-14 mm}
v_{\rm I}^{\rm d}(z_i,\theta_i) 
=[{\bi \mu}_i\cdot{\bar{\bi \mu}}_i-3 \mu_{zi}\bar{\mu}_{zi}]/2(2z_i)^3 
\nonumber\\
&&= -|{\bi \mu}_i|^2 (1+ \cos^2 \theta_i)/16z_i^3,
\ena  
where $\cos\theta_i= \mu_{zi}/|{\bi\mu}_i|$ with 
$\theta_i$ being the angle of ${\bi \mu}_i$ with 
respect to the $z$ axis. This   interaction energy  
is negative and grows for small $z_i$. 
In the vicinity of the wall, where 
  $|v_{\rm I}^{\rm d}|\gg k_BT$ holds,
dipoles are attracted   to the walls and are oriented 
in the  parallel direction ($\theta_i=0$) 
or in the antiparallel direction ($\theta_i=\pi$) 
with respect  to  the $z$ axis.

From the formulae in the previous section, 
we obtain the  counterparts 
by  replacing  $q_j$ by ${\bi \mu}_j\cdot\nabla_j$ for S-images
and $-q_j$ by ${\bar{\bi \mu}}_j\cdot\nabla_j$ for O-images, where 
$\nabla_j= \p/\p {\bi r}_j$. From  
 Eq.(2.21) the excess potential 
 $\phi({\bi r})$ for ${\bi r}\neq {\bi r}_j$ 
is obtained in the following superposition, 
 \be 
\phi =  \sum_{{\bi m}} {\sum_j}\nabla_j \cdot\bigg[ 
\frac{{\bi \mu}_j}{|{\bi r}-{\bi r}_{j} + 
{\bi h}|} +  
\frac{{\bar{\bi \mu}}_j}{|{\bi r}-{\bar{\bi r}}_{j}  
 + {\bi h}|} \bigg],
\en   
which surely vanishes at $z=0$ and $H$. 
The total electrostatic energy   in the fixed-potential condition 
 is written  as $U_{\rm m}^{\rm d}$.  
From   Eq.(2.23), we   find 
\bea 
&&\hspace{-1cm}  
U_{\rm m}^{\rm d}= \frac{1}{2} 
  \sum_{\bi m} \bigg[{\sum_{i,j}}'({\bi \mu}_i \cdot\nabla_i) 
({\bi \mu}_j \cdot\nabla_j) 
\frac{1}{|{\bi r}_{ij} +  {\bi h}|} \nonumber\\
&&\hspace{-1cm} + 
 {\sum_{i,j}}({{\bi \mu}}_i \cdot \nabla_i) 
({\bar{\bi \mu}}_j \cdot\nabla_j) 
\frac{1}{|{\bar{\bi r}}_{ij} +  {\bi h}|}\bigg]- E_a \sum_i \mu_{i z},  
\ena 
where   the self-interaction terms are removed in 
$\sum_{i,j}'$ and 
${\bi h}$ is given by   Eq.(2.22).   
The direct image interaction energy in Eq.(3.7) is included 
in the above expression.    
From  Eq.(3.1) the   differential form of 
$ U_{\rm m}^{\rm d} $ is given by    
\be 
d U_{\rm m}^{\rm d} 
= -\sum_i  ({\bi F}_i^e \cdot d{\bi r}_i  + 
  {\bi E}_i \cdot d{\bi \mu}_i)- \sum_i \mu_{ zi}   dE_a.
\en

The Ewald representation of $ U_{\rm m}^{\rm d}$  
may be written as  \cite{Klapp} 
\bea 
 U_{\rm m}^{\rm d}&=& \frac{1}{2}
\sum_{\bi m}  {\sum_{i,j}}' 
{\bi \mu}_i \cdot  {\aw}_s ({|{\bi r}_{ij} + {\bi h}|}) \cdot{{\bi\mu}}_j 
\nonumber\\
&&\hspace{-3mm}+\frac{1}{2}
\sum_{\bi m}  {\sum_{i,j}} 
{\bi \mu}_i \cdot  {\aw}_s ({|{\bar{\bi r}}_{ij} + {\bi h}|}) 
\cdot{\bar{\bi\mu}}_j  - \sum_i \frac{2\gamma^3 }{3\sqrt{\pi}} |{\bi \mu_i}|^2 
\nonumber\\
&&\hspace{-10mm} + \frac{1}{HL^2}  
 {\sum_{{\bi \nu}_\perp\neq{\bi 0}}}\sum_{i,j} \frac{\Psi_\ell(k)}{4}  
({\bi k}\cdot{\bi \mu}_i)  
{\bi k}\cdot(e^{{\rm i}{\bi k}\cdot{\bi r}_{ij}} {\bi \mu}_j  
+ e^{{\rm i}{\bi k}\cdot{\bar{\bi r}}_{ij}} 
{\bar{\bi \mu}}_j) 
\nonumber\\
&& +V_{\rm m}^{\rm d }  - E_a \sum_i \mu_{zi} .
\ena 
The first and second terms represent the short-range part with 
$\aw_s$ given in Eq.(3.4). The third term has  appeared 
 in  Eq.(3.3).  In the fourth long-range term, 
the summation is over ${\bi \nu}= (\nu_x,\nu_y,\nu_z)$  
with   ${\bi \nu}_\perp= 
(\nu_x,\nu_y)\neq {\bi 0}$ and ${\bi k}$ is given 
by Eq.(2.27). 
The fifth  term $V_{\rm m}^{\rm d}$  arises from the 
contributions with  $k_x=k_y=0$. From  Eqs.(2.26), (2.28),  and (C4), we find   \bea 
&&V_{\rm m}^{\rm d}=
 \frac{2\pi}{L^2} \sum_{ij} \mu_{zi}\mu_{zj}
 \bigg[\hat{\varphi}(z_i-z_j) 
+\hat{\varphi}(z_i+z_j)- \frac{1}{H}\bigg]\nonumber\\ 
&&\hspace{6mm} = \frac{4\pi}{HL^2}\sum_{n\ge 1}
 e^{-(\pi n/2\gamma H)^2} K_n^2,
\ena 
where  $K_n$ depends on $z_1, z_2, \cdots$ as 
\be 
K_n= \sum_j \mu_{zj}   \cos(\frac{\pi nz_j}{H}).
\en

The electric field ${\bi E}_i$ on dipole $i$ can be 
calculated from   Eq.(3.11).  That is, 
 the  contribution 
 from the first and second terms, 
that  from the third  and fourth  terms, 
and that from the  fifth and sixth    terms are 
written as ${\bi E}_{i}^{\rm ds}$, 
${\bi E}_{i}^{{\rm d}\ell}$, and 
 $E_i^{{\rm d}0 }{\bi e}_z$, respectively.  Then, 
\be  
{\bi E}_i = {\bi E}_{i}^{\rm ds}+ {\bi E}_{i}^{{\rm d}\ell }+ 
{E}_{i}^{{\rm d}0}{\bi e}_z .
\en 
From Eq.(3.12) ${E}_{i}^{{\rm d}0}$
 is expressed as  
\bea 
&& \hspace{-7mm} E_i^{d0}= E_a - \frac{4\pi}{L^2} \sum_{j}\mu_{zj}
\bigg[{\hat\varphi}(z_i-z_j)+
{\hat\varphi}(z_i+z_j)-\frac{1}{H}\bigg]  \nonumber\\
&&\hspace{-5mm}
=E_a-  \frac{8\pi}{HL^2}\sum_{n\ge 1}e^{- ({\pi n}/{2\gamma H})^2} 
 K_n \cos(\frac{\pi n z_i}{H}).
\ena 
In the first line, if the particles $j$ ($\neq i$) are 
randomly distributed in the cell, the second term  
vanishes, and the self term with $j=i$ is of order 
$\mu_{zi}\gamma/L^2$. Thus, if the boundary disturbances 
do not extend into the interior  
in a thick and wide  cell, we have $E_i^{d0}\cong  E_a$ (see Table 1 below). 
The first line of Eq.(3.15)   is consistent with  Eq.(D5) in Appendix D.


By replacing $q_j$ by $  {\bi \mu}_j\cdot\nabla_j$ in Eq.(2.30), 
 we obtain the surface charge densities, $\sigma_0(x,y)$ at $z=0$ and 
$\sigma_H(x,y)$ at $z=H$, in the   2D Fourier expansions,  
\bea 
&&\hspace{-4mm}\sigma_0=\frac{{E_a}}{4\pi} - 
 \frac{1}{L^2}\sum_{{\bi \nu}_\perp}\sum_j {\bi\mu}_j\cdot\nabla_j  
\frac{\sinh(kH-kz_j)}{\sinh(kH)} 
   e^{{\rm i}{\bi k}\cdot({\bi r}- 
{\bi r}_j)} ,\nonumber\\
&&\hspace{-4mm}\sigma_H=  - \frac{{E_a}}{4\pi} - 
 \frac{1}{L^2}\sum_{{\bi \nu}_\perp}\sum_j {\bi\mu}_j\cdot\nabla_j   
\frac{\sinh(kz_j)}{\sinh(kH)}  e^{{\rm i}{\bi k}\cdot({\bi r}- 
{\bi r}_j)},\nonumber\\   
\ena 
where ${\bi \nu}_\perp=(\nu_x,\nu_y)$ 
and  ${\bi k}=(2\pi \nu_x/L, 2\pi\nu_y/L,0)$. 
Integrating $\sigma_0$ and $\sigma_H$ 
in the planar region $0<x,y<L$ 
yields the total electric charges    
on the metallic surfaces,  
\be 
Q_0= -Q_H= \frac{1}{4\pi}L^2 E_a+ \frac{1}{H}\sum_j \mu_{zj}.
\en  
See Eq.(D6)  for   the counterpart 
in the continuum theory.

\subsection{Numerical example for dipoles between metallic plates under applied electric field}
 
\subsubsection{Model and method}

Next, we  present results   of 
molecular dynamics simulation of  
one-component spherical dipoles  
with number $N=1000$ between parallel metallic 
plates with $H=L$ at fixed potential difference $\Delta\Phi$.  

 As in Eq.(2.35),   
the total potential energy  is given by 
\be 
U_{\rm tot}= U_{\rm m}^{\rm d}  
+ \sum_{i>j}v_s(r_{ij})  +\sum_i v_{\rm w}(z_i),
\en  
where   $U_{\rm m}^{\rm d}$ is given by   Eq.(3.11),  
 $v_s(r)$ by Eq.(2.36), and $ v_{\rm w}(z)$ by Eq.(2.37).  
Hereafter, in terms of $\sigma$ and $\epsilon$ in  $v_s(r)$, 
units of length, density, electrostatic 
 potential, electric field, 
electric charge,  dipole moment, and temperature 
  are $\sigma$, $\sigma^{-3}$, $(\epsilon/\sigma)^{1/2}$, 
 $(\epsilon/\sigma^3)^{1/2}$, 
$(\epsilon\sigma)^{1/2}$,  $(\epsilon\sigma^3)^{1/2}$, and $\epsilon/k_B$, 
respectively. 
The average density  $n_0 = N / HL^2$ 
is chosen to be $0.19$, $0.38,$ and  0.57; then, 
$L= 17.4, 13.8$, and 12.0, respectively.

We cut off  $v_s(r)$ and  the short-range part of  
$ U_{\rm m}^{\rm d}$   in  Eq.(3.11)  
at  $r=r_{\rm cut}^s=4$. In $U_{\rm m}^{\rm d}$, we 
set  $\gamma=0.65$ and 
sum over $\bi k$  
 in the region  $k\le k_c=18\pi/L$. 
Thus, in $E_i^{{\rm d}0}$  in Eq.(3.15), the terms up to $n=18$ are 
summed. In addition, we note that 
 the simulation results are rather insensitive to 
the choice of  $\gamma$. 
To confirm  this, we also performed  simulations 
for $\gamma =0.5$ and  $1$ with 
$r_{\rm cut}^s=6$ for the case of 
 $n_0=0.19$, $T=1.0$, and $E_a=0.058$. 
Then  the numbers in the first line of Table 1 
were changed only slightly with differences less than  $5\%$.


The  dipole moment   has  a fixed 
magnitude   $\mu_0$ as 
\be 
{\bi \mu}_i= \mu_0 {\bi n}_i,
\en 
where ${\bi n}_i$ is the  unit vector 
representing  the direction of polarization.  
In this subsection, we set  
\be  
\mu_0=|{\bi \mu}_i|= 2 (\epsilon\sigma^3)^{1/2}.
\en 
Then our system is 
highly polarizable 
at high densities. 

The total kinetic energy is assumed to be of  the form, 
\be 
K_{\rm tot}=   \sum_i \frac{m}{2} |{\dot{\bi r}}_i|^2
+\sum_i \frac{I_1}{2} |{\dot{\bi n}}_i |^2,
\en 
where ${\dot{\bi r}}_i=d {\bi r}_i/d t$, ${\dot{\bi n}}_i= 
d {\bi n}_i/d t$, and 
$I_1$  is the moment of inertia. 
Regarding dipoles as spheres with diameter $\sigma$, 
we set  $I_1= m\sigma^2/10$. The Newton  equations  of motions 
for ${\bi r}_i$ and ${\bi n}_i$  are written as\cite{EPL}   
\bea
&& \hspace{-4mm} m \frac{d^2}{d t^2} 
{{\bi r}}_i =-\pp{}{{\bi r}_i} U_{\rm tot}, \\ 
&&\hspace{-10mm}  I_1  {\bi n}_i \times  \frac{d^2}{d t^2} {{\bi n}}_i=
-{\bi \mu}_i \times \pp{}{{\bi \mu}_i}U_{\rm tot},  
\ena  
where ${\bi n}_i\cdot{d^2{\bi n}_i/dt^2}= 
-|{\dot{\bi n}}_i|^2$ from ${\bi n}_i\cdot{\dot {\bi n}}_i=0$. 
The total energy $U_{\rm tot}+K_{\rm tot}$ is  conserved in time. 
 We integrated these equations in the NVE ensemble using 
the leap-frog method with the time step width being 
$0.002$, as in the previous section (see  
 below Eq.(2.40)). The translational 
and rotational kinetic energies 
were kept at $3k_BT/2$ and  $k_BT$, 
respectively,  per particle.

\subsubsection{Highly polarizable liquids}

\begin{figure}
\includegraphics[width=0.96\linewidth]{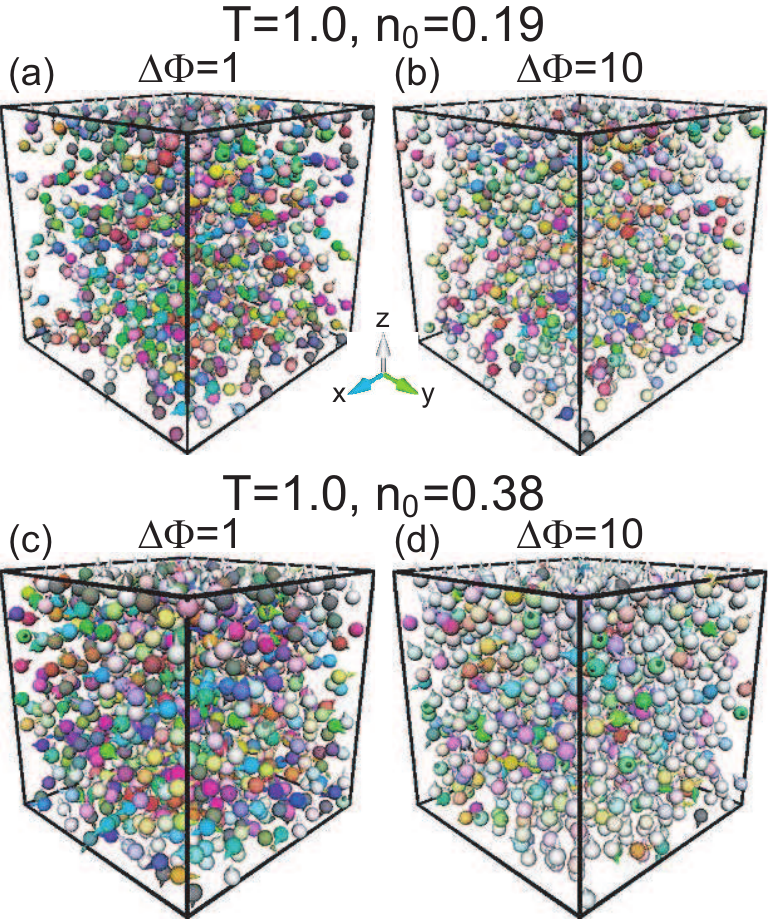}
\caption{(Color online) Snapshots of spherical 
dipoles   in liquid at $T=1$:
(a) $n_0=0.19$ and   $\Delta \Phi=1$, 
(b) $n_0=0.19$ and   $\Delta \Phi=10$, 
(c) $n_0=0.38$ and   $\Delta \Phi=1$, 
and (d) $n_0=0.38$ and   $\Delta \Phi=10$. 
Colors  represent the orientation of ${\bi \mu}_i$ as in Fig.7.   
}
\end{figure}

\begin{figure}[h]
\includegraphics[width=1\linewidth]{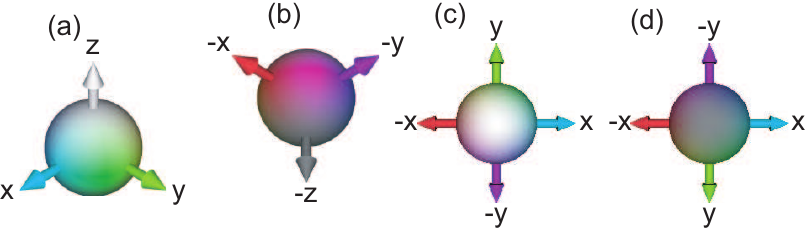}
\caption{(Color online) 
Color maps of polarization direction ${\bi n}_i$ 
on a sphere surface: (a) Diagonally downward view, 
(b) diagonally upward view, 
(c) top view, 
and (d) bottom  view.  
Spheres  become  white (gray) 
as they are parallel (antiparallel) to the 
applied field. 
}
\end{figure}  
\begin{figure}
\includegraphics[width=0.96\linewidth]{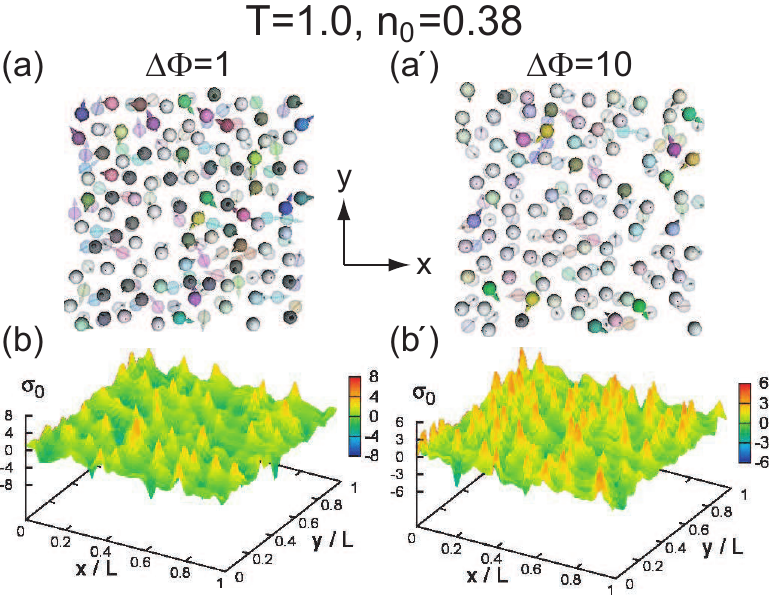}
\caption{(Color online) 
Top: Dipole  configurations  
in the two layers  $0<z<1$  and $1<z<2$ for   
$\Delta \Phi=1$ (left) and  $10$ (right), 
 according to the color maps in Fig.7.   
Bottom: Surface charge density 
$\Delta\sigma_0(x,y)$ 
at $z=0$ for  $\Delta \Phi=1$  and $10$  
in units of $(\epsilon/\sigma^3)^{1/2}$. 
Data are common to those  in (a) and (a$'$) of  Fig.6. 
}
\end{figure} 
\begin{figure}
\includegraphics[width=0.96\linewidth]{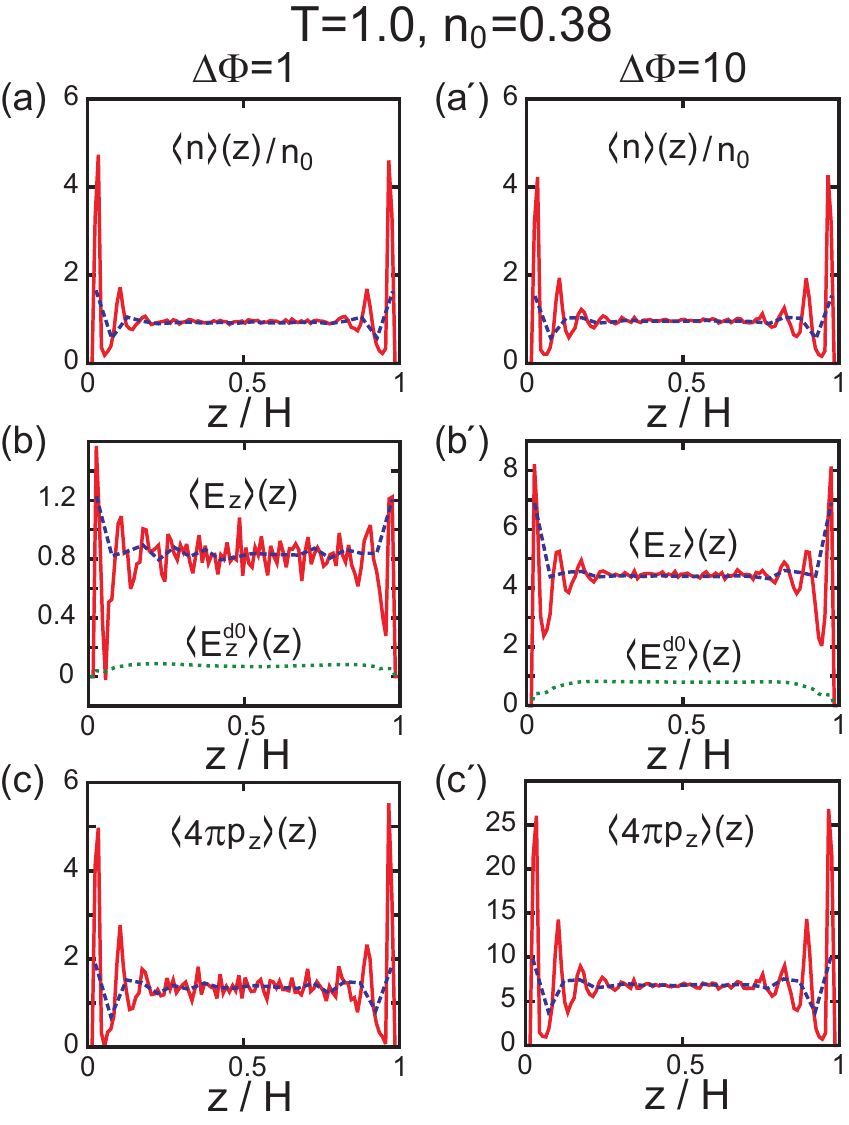}
\caption{(Color online) Simulation results of spherical 
dipoles  for $n_0=0.38$ and $T=1$ in liquid, where 
$\Delta \Phi=1$ (left) and  $10$ (right). 
Normalized lateral density average $\av{n}(z)/n_0$ is in (b) and (b$'$). 
Lateral average $\av{E_{z}}(z)$ are in (c) and (c$'$) 
and that  $\av{4\pi p_{z}}(z)$ are in (d) and (d$'$). Here, 
$\Delta z=H/100$ (bold red  lines) 
and $\Delta z=H/20$ (broken blue lines), where $\Delta z$ is the 
smoothing length. 
}
\end{figure}

In  Fig.6,  we show snapshots of spherical dipoles 
  for $n_0=0.19$ and $0.38$    in liquid  at 
$T=1$ under   $\Delta\Phi=1$ and 10. 
The colors of the spheres  represent 
 their polarization directions according to 
the color maps in Fig.7. We can see  
orientation enhancement with increasing $\Delta\Phi$. 
For $n_0=0.19$,   chainlike 
associations are clearly visualized, whose lengths 
increase with increasing     $\Delta\Phi$. 
For  small  $\Delta\Phi$ a large number of dimers 
 appear.
 
In the top panels of  Fig.8, we   display  
particle configurations near the 
bottom wall in the two layers, $0<z<1$ and  $1<z<2$,  
and the surface charge density $\sigma_0(x,y)$ 
 in Eq.(3.16).  In these cases, dipoles are accumulated 
on the walls and are oriented. 
For $\Delta\Phi=1$,    
the  image interaction (3.7) is strong 
enough to induce   alignment both  in the 
parallel and antiparallel directions 
along  the $z$ axis. 
For $\Delta\Phi=10$ those in the parallel 
direction  increase.  
In the bottom 
panels of Fig.8,  the surface  charge density 
in Eq.(3.16) are written, which are highly 
 heterogeneous 
and fluctuating in time.

In Figs.9(a)-(a$'$), 
we display   the 
laterally averaged density $\av{n}(z)$ 
defined as in Eq.(2.41), where we write  two curves 
for $\Delta z=H/100=0.138$ 
and $H/20=0.69$. 
We recognize that  the interior away from the walls is 
in a homogeneous, uniaxial equilibrium 
state  under electric field for $H\gg 1$. 
In Table 1,  the bulk density $\bar{n}$ in the interior 
is slightly smaller than $n_0$ 
because  the dipoles are accumulated  near the walls. 
We  determined  the interior (bulk) 
values from the lateral averages 
 with  $\Delta z=H/20$.


\begin{table}
\caption{Data  for 1000 spherical dipoles 
in liquid at $T=1$ in a $L\times L\times H$ cell   
($H=L$) under applied potential difference 
$\Delta\Phi=HE_a= 1$ and 10  for three  
densities $n_0=0.19$, $0.38$, and $0.57$.
Here,  $\bar n$, ${\bar\mu}_z$, and 
$ {\bar p}_z={\bar n} {\bar\mu}_z$ 
are  bulk values in the interior. 
Also given are  bulk values of $\av{E^{{\rm d}0}}(z)$ 
and  $\av{E_z}(z)$, dielectric constant $\varepsilon_{\rm e}$ 
in Eq.(3.29), local field factor  
 $\gamma_{\ell}$ in Eq.(3.30), polarizability $\alpha_{\rm e}= {\bar \mu}_z/
\av{E_z}$. 
}  
\begin{tabular}{|cc|ccccccccc|}
\hline
$n_0$  &
$E_a$&
$Q_0$&
$\av{E^{{\rm d}0}}$&
${\bar{n}}/{n_0}$&
${\bar \mu}_z$&
$4\pi {\bar p}_z$  &
$\av{E_z}$&  $\varepsilon_{\rm e}$ &
$\gamma_{\ell}$&
$\alpha_{\rm e}$\\
\hline
0.19 &0.058& 11.6 &0.060&0.88& 0.18 &0.37&0.42 &7.4&0.97&0.43 \\
\hline
0.19 &0.58& 78.5 &0.65&0.91& 1.20 &2.62&2.88&5.5&0.88&0.42 \\
\hline
0.38 & 0.073&23.0&0.069&0.95&0.30 &1.36&0.82&19.6&0.55&0.37 \\
\hline
0.38 & 0.73&115&0.80&0.97&1.46 &6.81&4.45&10.3&0.55&0.33 \\
\hline
0.57 &0.083&40.5&0.069&0.97& 0.48&3.29&1.55 &40.6 &0.45&0.31\\
\hline
0.57 &0.83&143&0.95&0.98& 1.62&11.36&5.89&14.7 &0.45&0.26  \\
\hline
\end{tabular}
\end{table}

Using the decomposition (3.14) 
of the local field ${\bi E}_i$, we 
define  the  laterally averaged 
 electric fields   $\av{E_z}(z)$,  
 $\av{E_z^{\rm ds} }(z)$, 
$ \av{E_z^{{\rm d}\ell} }(z)$, 
and $ \av{ E^{{\rm d}0}}(z)$  as 
in Eq.(2.42). Hereafter,  $\av{\cdots}$ represents 
the average over a long time  interval 
with  width $10^3$. 
Along the $z$ axis we thus have 
\be 
\av{E_z}(z)=  \av{E_z^{\rm ds} }(z)+  \av{E_z^{{\rm d}\ell} }(z)+ 
\av{E^{{\rm d}0}}(z). 
\en 
If the boundary layers are 
much thinner than $H$, 
we have  
$ \av{E^{{\rm d}0}}(z)\cong  E_a$  in the interior (see 
the sentences below Eq.(3.15)).
For the examples in   Table 1, we can indeed see 
  $ \av{E^{{\rm d}0}}\cong E_a$. 
We also find that 
$\av{E_z^{{\rm d}\ell} }(z)$ is of the same 
order as $E_a$ \cite{ratio}.
In the bulk region of our examples, 
the short-range contribution 
$\av{E_z^{\rm ds} }(z)$ is thus 
dominant in the right hand side of 
Eq.(3.24), typically being about 
 $90\%$ of $\av{E_z }$, so that  
\be 
\av{E_z} \cong \av{E_z^{\rm ds} }. 
\en  
We also define the lateral average of the polarization, 
\be 
\av{p_z}(z)= 
\sum_j\av{ \mu_{zj} \frac{\theta(z_j-z)\theta(z+\Delta z-z_j)}{L^2\Delta z}}.   \en 
We plot $\av{E_z}(z)$ in Fig.9(b)-(b$'$) 
and $\av{p_z}(z)$ in  Fig.9(c)-(c$'$)  
 for $\Delta z=H/100$ and $H/20$, where the 
curves are more smooth  for larger  $\Delta z$. 
In the interior, they assume  
the bulk average values with small fluctuations.  
In Table 1, we give the average bulk values 
of $\av{E_z}$, $\av{E^0}(z)$, and $\av{4\pi E_z}(z)$ at $T=1$ 
for three densities $n_0=0.19, 0.38$, and 0.57 under $\Delta\Phi=1$ and 10.

 In terms of  
 the polarization variable,   
\be 
\hat{\bi p}({\bi r})= \sum_i {\bi \mu}_i \delta ({\bi r}_i -{\bi r}),
\en  
we have 
$\av{p_z}(z)= \int_0^L dx \int_0^L dy \int_{z}^{z+{\Delta z}}dz 
{\hat{p}}_z({\bi r})
/L^2\Delta z$. Averaging  
 over the  particles in the interior  and over a long time yields 
the  bulk  average polarization,   
\be 
{\bar p}_z = \av{\hat{p}_z}=  \bar{n} {\bar\mu}_z, 
\en 
where  ${\bar\mu}_z$ is the  average of $\mu_{zi}$ in the interior. 
We define  the effective dielectric constant   by 
\be 
\varepsilon_{\rm e} = 1+ 4\pi {\bar p}_z/ E_a. 
\en 
In Table 1,  $\varepsilon_{\rm e}$  is much  
larger than unity, 
increasing  with increasing $n_0$ and decreasing  
with increasing $E_a$.  For $\Delta\Phi=10$, 
 ${\bar\mu}_z$ approaches 
its  maximum  $\mu_0=2$, where the system is in the 
nonlinear response regime.  In addition, 
for $n_0=0.57$ and $\Delta\Phi=1$, 
$\varepsilon_{\rm e}$ is given by a large value of 
40. With lowering $T$ at this density, we find occurrence of 
a  ferroelectric phase transition 
 around $T\cong 0.5$, on which we will report shortly.

Furthermore, 
since ${\bi E}_i$ is the local electric field, 
its lateral average $\av{E_z}$ along the $z$ axis 
is related to the applied electric field $E_a$  and the  
local polarization ${\bar p}_z$ by 
\be 
\av{E_z}= E_a + 4\pi  \gamma_{\ell} {\bar p}_z, 
\en 
in the bulk region. The  second term represents  the   
 Lorentz field  with $\gamma_{\ell}$ being 
 the local field factor\cite{Onsager,Kirk,local}.
The  classical value of $\gamma_{\ell}$  is $1/3$. However, in 
 Table 1,  $\gamma_{\ell}$ considerably exceeds  
 $1/3$. It  increases with decreasing $n_0$  
but  is rather insensitive to $E_a$. 
We may also introduce 
 the  polarizability 
$\alpha_{\rm e}={\bar\mu}_z/\av{E_z}$. 
Then, $\varepsilon_{\rm e}$ satisfies   
\be 
\frac{{\varepsilon_{\rm e}}-1}{{1+\gamma_\ell (\varepsilon_{\rm e}-1)}  
}= 4\pi {\bar n} \alpha_{\rm e},
\en  
which nicely holds in Table 1. The  
Clausius-Mossotti formula follows for $\gamma_\ell=1/3$.
Here, we should  suppose 
 a thick cell ($H \gg 1$) 
 to avoid the boundary effect in 
the relations (3.29) and (3.30), 
though our system size is still 
not large enough.

In our case,   the classical local field 
relation breaks down because of strong   pair correlations 
along the $z$ axis. To explain this, 
we assume   a large homogeneous interior  
region in liquid for large $H$ and $L$. 
There, we  define the pair correlation functions 
$g(r,\theta)$ and $g_{\rm p}(r,\theta)$  by 
\bea 
&&\hspace{-12mm} \av{\hat{n}({\bi r_0}) \hat{n}({\bi r_0}+{\bi r}) }
= {\bar n} \delta ({\bi r}) + {\bar n}^2 g(r, \theta) ,
\\   
&&\hspace{-12mm} \av{\hat{n}({\bi r_0}) \hat{{p}}_z({\bi r_0}+{\bi r}) }
= {\bar n}\bar{{ p}}_z \delta({\bi r}) +  
 {\bar n} { g}_{\rm p}(r, \theta), 
\ena 
where  
$\hat{n}({\bi r}) =  \sum_i \delta ({\bi r}_i -{\bi r}_0)$ 
is   the density variable   and  
$\hat{\bi p}({\bi r})$ is defined in Eq.(3.27). 
These relations  are independent 
of ${\bi r}_0$ from the translational invariance.
The $g(r,\theta)$ and 
$g_{\rm p}(r,\theta)$ depend only on $r$ and $ \theta=\cos^{-1}(z/r)$ 
from the rotational invariance around the $z$ axis. 
For $r \gg 1$,   we have   $g(r,\theta)\to 1$ and 
$g_{\rm p}(r,\theta)\to {\bar p}_z$.  
Also from the 
invariance with respect to the inversion ${\bi r} 
\to -{\bi r}$,
they are even functions of $\pi/2- \theta$ and   we notice 
 \be 
\av{\hat{n}({\bi r_0}) \hat{{p}}_x({\bi r_0}+{\bi r}) }
= \av{\hat{n}({\bi r_0}) \hat{{p}}_y({\bi r_0}+{\bi r}) }=0.
\en 
In Fig.10, we plot $g(r,\theta)$ and 
$g_{\rm p}(r,\theta)/{\bar p}_z$ 
for $T=1$, $n_0=0.38$, and $\Delta\Phi=10$. 
They exhibit first peaks at $r \sim 1.1$ 
and second peaks at $r \sim 2.2$. 
They  are maximized for $\cos\theta =\pm 1$, 
so the associated dipoles 
are on the average oriented along the $z$ axis. 
However, this tendency   is not clearly visualized   
in Fig.6(d) at $n_0=0.38$, 
while it is evident in Fig.6(b) at  $n_0=0.19$. 
For ferromagnetic fluids, 
similar behavior of the pair correlation 
function $g(r,\theta)$ 
was theoretically examined  \cite{PG} 
and numerically calculated 
 \cite{Ferro}.

\begin{figure}[t]
\includegraphics[width=0.96\linewidth]{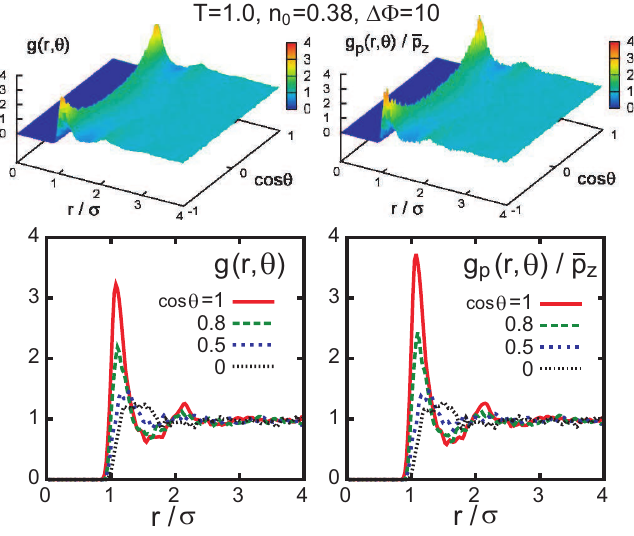}
\caption{(Color online) Pair correlation functions 
$g(r,\theta)$ (left) 
and $g_{\rm p}(r,\theta)/{\bar p}_z$ (right)  
of  dipoles in liquid 
 on  the $r$-$\cos\theta$ plane (top) 
and   as functions of $r$ 
for $\cos\theta= 1, 0.8$, 0.5, and 0 (bottom), 
where  $n_0=0.38$,  $T=1$, and $\Delta\Phi=10$. 
They exhibit  peaks at $r \sim 1.1$ 
and at $r \sim 2.2$ along the field  ($\cos\theta \cong \pm 1$). 
}
\end{figure} 

From   the first term in  Eq.(3.11),   
 the short-range 
part    of the local field in the interior 
is given by  
${\bi E}_i^{\rm ds}\cong -\sum_{j\neq i} 
\aw_s(|{\bi r}_{ij}|)\cdot {\bi \mu}_j $,  
where the contributions  with ${\bi m}\neq 
(0,0,0)$ are neglected.  From Eqs.(3.4) and (3.33) 
 the  lateral average of its $z$ component is 
written in terms of $g_{\rm p}(r,\theta)$ for $\Delta z\ll 1$ as 
\be
\av{E_z^{\rm ds}}\cong 
 \int d{\bi r} [C(r) \cos^2\theta-B(r)]g_{\rm p}(r,\theta). 
\en  
  If  we set 
$g_{\rm p}(r,\theta)= {\bar p}_z$, 
the integral is nearly equal to 
 $4\pi {\bar p}_z/3$ from Eq.(3.5) 
\cite{Lorentz}. Use of  Eq.(3.25)  yields  
the correction to the classical value in the form,   
\be
\gamma_{\rm \ell}
- \frac{1}{3}  \cong  
 \int \hspace{-1mm}
 \frac{d{\bi r}}{4\pi}
 [C(r) \cos^2\theta-B(r)]
[\frac{g_{\rm p} (r,\theta)}{{\bar p}_z}-1] .
\en 
Here,   we may well replace 
$C(r) \cos^2\theta-B(r)$ 
by $(3\cos^2\theta-1)/r^3$, since 
${g_{\rm p} (r,\theta)}/{{\bar p}_z}-1$ 
is small for $r/\sigma \gs  \gamma^{-1}$ 
in Fig.10.  In Table 1, 
the above formula reproduces  
 the numerical values of $\gamma_\ell$  
with errors   of order   $10\%$.

\subsubsection{Dipole chains} 

\begin{figure}[t]
\includegraphics[width=0.96\linewidth]{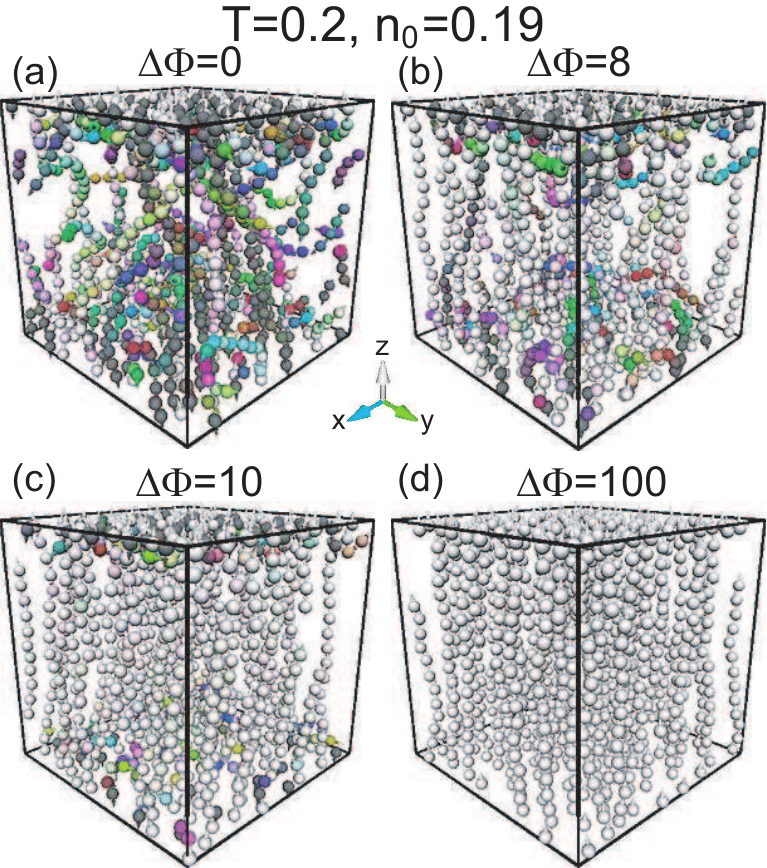}
\caption{(Color online) 
 Chains of 1000 spherical dipoles  at $T=0.2$ for $n_0=0.19$  
 in a $L\times L\times H$ cell ($H=L$). 
Potential difference $\Delta\Phi$ 
is set equal to 0, 8, 10, and 100.  
For small $\Delta\Phi$ dipoles parallel and 
antiparallel to the $z$ axis are both 
attracted and aligned by 
the image interaction near the walls. 
They are written in white 
or in gray, respectively.   For large  $\Delta\Phi$,  
they are  aligned along the $z$ axis.}
\end{figure} 
\begin{figure}[t]
\includegraphics[width=0.96\linewidth]{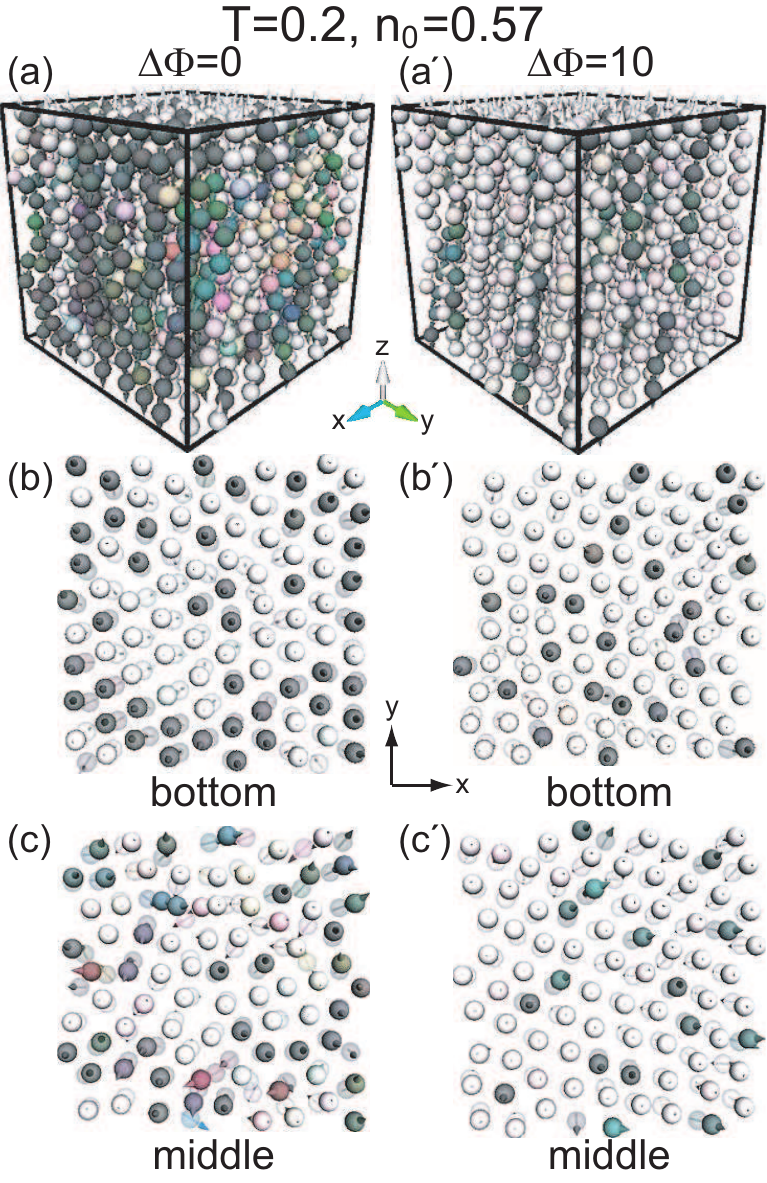}
\caption{(Color online) 
 Chains of 1000 spherical dipoles  at $T=0.2$ for $n_0=0.57$  
 in a $L\times L\times H$ cell ($H=L$) 
for $\Delta\Phi=0$ (left) and 10 (right).  
Shown also are cross-sectional configurations 
in  the bottom two layers,  $0<z<1$ and $1<z<2$,  
in (b) and (b$'$) 
and those in  the middle  two layers, $H/2+1<z<H/2+2$ and 
$H/2<z<H/2+1$, in (c) and (c$'$). 
White (gray) particles are parallel (antiparallel) 
to the $z$ axis.   Colors are lighter 
for dipoles  in the second layers.}
\end{figure} 
\begin{figure}[t]
\includegraphics[width=0.96\linewidth]{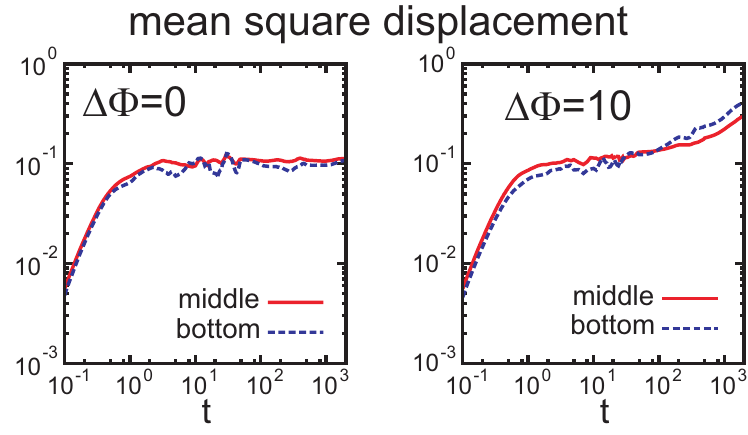}
\caption{(Color online) 
Mean square displacements 
for the particles near the walls and those 
far from them 
for $\Delta\Phi=0$ (left) and 10 (right), 
 where $T=0.2$ and  $n_0=0.57$. 
The data  are the same as  in Fig.12. 
Time $t$ is in units of $\tau_0$ in Eq.(2.40).  
}
\end{figure}

For ferromagnetic particles 
with  permanent dipoles,  
chain formation was 
predicted  \cite{PG} and has been studied numerically  
 \cite{Weis,Mazars,KlappJCP}.
As electrorheological fluids,  
 use has been made of colloids with a dielectric constant 
different from that of the surrounding fluid \cite{Hasley,Martin},  
which  have  induced dipole 
moments in  electric field.  
Coating  of colloids with  urea (with a high molecular 
dipole moment of 4.6D) is known to increase 
the polarizability \cite{Shen}.  
Experiments on such dielectric particles have been performed 
extensively, but 
simulations including the image interaction 
under  applied electric field 
were performed only for  induced  point dipoles 
oriented  along the $z$ axis\cite{Tao}.
Here, we present simulation 
results on   permanent point dipoles. 
We show  that they  are attracted to the walls 
by  the image dipoles  and the 
surface charges due to the applied electric field.

In  Fig.11, we  gives our  
examples of dipole chains at $T=0.2$ 
for  $n_0=0.19$ under  four potential differences 
$\Delta \Phi=0$, 8, 10, and 100. 
 For $\Delta\Phi= 0$ in (a), 
chains are formed  near the walls, 
where dipoles are attracted to the walls 
and aligned in the directions  
parallel or antiparallel to the $z$ axis  
due to the image interaction (3.7). 
For   $\Delta\Phi= 8$ in (b), even dipoles away from 
the walls form strings along the $z$ axis. 
 For $\Delta\Phi= 10$ in (c), 
chains are stretched  
between the two plates, but antiparallel 
wall attachments  still occur. 
For  $\Delta\Phi=100$ in (d), all the dipoles 
 are aligned along the $z$ axis.

We  consider the average dipole moment 
over all the particles and that over those 
at  the bottom defined by     
\be 
\av{\mu_z}_{\rm all} = \sum_i \mu_{zi}/N, \quad 
\av{\mu_z}_{\rm wall} = \sum_{0<z_i<1} \mu_{zi}/N_{\rm w}, 
\en  
where $N_{\rm w}$ is the particle number in the 
layer  $0<z_i<1$. 
For the data in Fig.11,  we obtain 
$(\av{\mu_z}_{\rm all},\av{\mu_z}_{\rm wall})= 
(0,0), (0.67,0.09), (1.34, 0.27),(1.91,0.69), 
(1.97,1.97)$ 
for $\Delta \Phi=0$, 8, 10, 100, respectively. 
Here,  we estimate  the image interaction 
energy  as 
$|v_{\rm I}^{\rm d}| \sim 4$ 
for $z_i \sim 0.5$ from Eq.(3.7) 
and  the typical field  energy as  $\mu_0 E_a  
\sim 0.2 \Delta\Phi$. Comparison  
of these two energies yields 
a  crossover potential difference 
 about 20, in accord   with Fig.11.

In Fig.12, we show  dipole configurations 
 for $n_0=0.57$ under two  potential differences 
 $\Delta\Phi=0$ and 10. 
At this high density,  a hexagonal lattice (with defects) 
appears on the walls,   
 extending   between the top and the bottom 
 in the form of chains. 
For $\Delta\Phi=0$,  
the cross-sectional   configurations 
in the middle are glassy. We can see that the chains are 
somewhere curved, broken, branched, or  even entangled. 
In the left panel of Fig.13, 
we plot the mean-square displacement,  
\be   
M(t)= \av{|{\bi r}_i(t+t_0)-{\bi r}_i(t_0)|^2}, 
\en 
where the average over $t_0(<10^4)$ 
was  taken.  Furthermore, we took  the average 
over the particles close to the walls 
and that over those in the interior separately, 
but there was no essential difference 
for these two groups  of the particles, as shown 
in Fig.13. 
We notice  that $M(t)$ 
grows to saturate at  a plateau value ($\sim 0.1$) 
for $t \gs 1$  in units of $\tau_0$ in Eq.(2.40).  
We can  also see 
 a tendency  of  segregation   between 
 the chains   parallel to 
the $z$ axis  and those antiparallel to the $z$ axis.   
On the other hand, for $\Delta\Phi=10$,  
the fraction of the chains  parallel to   
the $z$ axis increases  
and the  orientational fluctuations  decrease, 
where $ \av{\mu_z}_{\rm all}=1.17$ 
and $ \av{\mu_z}_{\rm wall}=0.99$ (see Eq.(3.37)). 
In this more ordered state, 
collective configuration changes 
are   appreciable for $t \gs 100$. In the right panel of 
Fig.13, this leads 
 to very slow growth of $M(t)$.

\section{Summary and remarks}

In summary, we have extended the  Ewald method 
for   charged and polar particles between  metallic plates 
in a $ L\times L\times H$ cell,   
aiming to  apply electric field $E_a$ to these systems.  
In this problem, we should account for 
an infinite number of image  charges and   dipoles outside the cell. 
With this method, 
we have  presented 
some  results of molecular dynamics simulation.

In the previous papers  
 by   Hautman {\it et al.}\cite{Hautman}, 
by Perram and  Ratner\cite{Perram}, and by Klapp \cite{Klapp}, 
the conventional 3D Ewald method for periodic systems 
was applied to a doubly  expanded 
$ L\times L\times 2H$ cell, where the three axes 
were formally equivalent.  
Our scheme is essentially the same 
as  theirs, but  we have  treated   the $z$ direction  
differently from the lateral   directions.  
In particular,  we  have divided the terms in the long-range part of 
 the Ewald  sum into those inhomogeneous 
 in the $xy$ plane   (with nonvanishing $k_x$ or $k_y$) 
and those homogeneous in the $xy$ plane 
but inhomogeneous along  the $z$ axis (with $k_x=k_y=0$ and  $k_z\neq 0$) 
in Eqs.(2.26) and (3.11). 
The latter one-dimensional 
terms can be summed  into simple forms  in Eqs.(2.28) 
and (3.12),  yielding the one-dimensional electric field  $E_i^0$ 
in Eqs.(2.33) and (3.15).  In our simulations,  
we have calculated $E_i^0$ very accurately. 
For dipole systems, $E_i^0\cong E_a$ far from the walls 
for thick cells.

In applying electric field between metallic plates, 
we may control 
the potential difference 
$\Delta\Phi= H E_a$ or 
 the surface charge  $Q_0$. 
The electrostatic  energy, 
$U_{\rm m}$ or  $U_{\rm m}^{\rm d}$,  
in   the fixed-potential condition  
is  related to that 
in   the fixed-charge condition   
by the Legendre 
transformation, as shown in Appendix B. 
In the continuum electrostatics in Appendix D, 
we have also introduced the free energies  $F_{\rm m}$ and $F_{\rm e}$ 
for  these two cases \cite{Landau,OnukiNATO}.

Some remarks are given below.\\ 
(1) We have assumed  stationary 
applied  field, but  
we may assume  a time-dependent  field $E_a(t)$   
and examine  various nonequilibrium phenomena.
 For example, when $\Delta\Phi$ was changed from 0 to 
100, we observed complex dynamics of charged 
particles from the crystal in Fig.4(a) 
to that in Fig.4(a$'$). 
We also mention that melting due to 
electric field  was observed 
in charged colloidal  crystal \cite{super}.\\
(2) We have assumed spherical dipoles, 
but real molecules are nonspherical 
and undergo hindered rotations. 
This feature should be included in future simulations. 
We should also  consider 
mixtures of ions and nonspherical polar molecules 
bounded by metallic plates, 
where the ion-dipole interaction is crucial   
\cite{Is,Onuki}.\\ 
 (3) The classical result  $\gamma_\ell=1/3$ 
for the local field factor 
follows  for a spherical cavity \cite{Onsager,Kirk}, 
leading to the Clausius-Mossotti formula 
for the dielectric constant. However, 
$\gamma_\ell$  becomes one of the depolarization 
factors for an ellipsoidal cavity\cite{local}, so 
it is sensitive to the  environment 
around each  dipole. 
In our case, $\gamma_\ell$  increases due to the 
 pair correlations along the applied field 
as in Eq.(3.36). 
More systematic simulations are needed  
on this aspect.\\
(4)  Various systems such as 
charged colloids, polyelectrolytes, 
proteins,  and water molecules   
  should exhibit interesting behaviors  
close to  metal surfaces  without and with 
applied field \cite{colloid, Corni,Messina}. 
For polarizable surfaces, a simulation 
method similar to ours has recently been 
reported \cite{double}. 
\\
(5) We may well expect ferroelectric phase transitions 
in confined dipole systems (see the sentences below Eq.(3.29)). 
Furthermore, it is  of great interest  to examine 
the electric field effects in ferroelectric 
systems with impurities.
  We will shortly report on the   
electric field effect in orientational glass 
as a continuation of our previous work  \cite{EPL}.\\

\begin{acknowledgments}
This work was supported by Grant-in-Aid 
for Scientific Research  from the Ministry of Education, 
Culture,  Sports, Science and Technology of Japan. 
K. T. was supported by the Japan Society for Promotion of Science.
The numerical calculations were carried out on SR16000
at YITP in Kyoto University.
\end{acknowledgments}

\vskip5mm
\hspace{-0.3cm}
\noindent 
{\bf Appendix A: Derivation  of  Eq.(2.7)}\\
\setcounter{equation}{0}
\renewcommand{\theequation}{A\arabic{equation}}

The last two terms in  Eq.(2.7) arise from  the long-range part of 
the electrostatic energy, written as $ U_{\rm p}^\ell$. 
We  transform it as follows: 
\bea 
 U_{\rm p}^\ell &=& \frac{1}{2} 
\sum_{\bi m}  {\sum_{i,j}}  
{q_iq_j}\psi_\ell ({|{\bi r}_{ij} + L{\bi m}|})\nonumber\\
&=& \frac{1}{2} \int_{\bi k} \Psi_\ell(k) 
\sum_{\bi m}  {\sum_{i,j}}  
{q_iq_j} e^{{\rm i}{\bi k}\cdot ({{\bi r}_{ij} + L{\bi m}})}, 
\ena 
where  $\int_{\bi k}= (2\pi)^{-3} \int d{\bi k}$ and 
$\Psi_\ell(k)$ is defined in Eq.(2.8). 
In the second line, we    perform the summation 
over ${\bi m}$  introducing a damping 
factor $\exp[{-\epsilon |m_x|-\epsilon |m_y|- \epsilon|m_z|}]$, 
where  $\epsilon$ is a positive small  number (not 
to be confused with the energy $\epsilon $ in $v_s(r)$ in Eq.(2.36)). 
The summation yields $\sum_{\bi m}\exp({{\rm i}{\bi k}\cdot  L{\bi m}})= 
\zeta(Lk_x )\zeta(Lk_y )\zeta(Lk_z )$, where 
\bea 
\zeta(u )&=& \sum_{m_x=0,\pm 1, \cdots} \exp[ {\rm i} 
m_x u - \epsilon |m_x|) 
\nonumber\\&&\cong {2\epsilon}/[{\epsilon^2+ 2(1-\cos(u))}].
\ena 
For $L k >1$, $\Psi_\ell(k)$ 
is finite, so  we may  replace 
$\zeta(Lk_\alpha)$ by $ 2\pi\sum_{\nu_\alpha \neq 0} 
\delta (Lk_\alpha-2\pi \nu_\alpha)$, where $\alpha=x,y,z$. 
For  $ Lk\ls  1$, we may set   
$\zeta(Lk_\alpha)=  {2\epsilon}/[{\epsilon^2+ L^2k_\alpha^2}]$. 
Thus, $U_{\rm p}^\ell$ becomes  
\bea 
 U_{\rm p}^\ell&=& \frac{1}{2} \int_{\bi k} \Psi_\ell(k) 
\zeta(Lk_x) \zeta(Lk_y) \zeta(Lk_z) 
  {\sum_{i,j}}  
{q_iq_j} e^{{\rm i}{\bi k}\cdot {\bi r}_{ij}} \nonumber\\
&=&  \frac{1}{2L^3} \sum_{{\bi k}\neq {\bi 0}} \Psi_\ell(k) 
 {\sum_{i,j}}  
{q_iq_j} e^{{\rm i}{\bi k}\cdot {\bi r}_{ij}} \nonumber\\
&&\hspace{-8mm} +\int_{\bi k}  \frac{16\pi\epsilon^3 
 {\sum_{i,j}}  
{q_iq_j} [ e^{{\rm i}{\bi k}\cdot{{\bi r}_{i}}}-1] 
 [ e^{-{\rm i}{\bi k}\cdot{{\bi r}_{j}}}-1] 
}{ k^2({\epsilon^2+ L^2k_x^2})( 
{\epsilon^2+ L^2k_y^2})({\epsilon^2+ L^2k_z^2})} .
\ena 
The first term   coincides with  the third term in Eq.(2.7) 
 with ${\bi k}= 2\pi L^{-1}(\nu_x, \nu_y, \nu_z)\neq {(0,0, 0)}$. 
The second term arises  from  $k\ls L^{-1}$, 
where  
$ \exp[{{\rm i}{\bi k}\cdot {\bi r}_{i}}]- 1 \cong 
{{\rm i}{\bi k}\cdot {\bi r}_{i}}$ 
and 
$ \exp[-{{\rm i}{\bi k}\cdot {\bi r}_{j}}]- 1 \cong 
-{{\rm i}{\bi k}\cdot {\bi r}_{j}}$  for small $k$.
Furthermore, 
 ${k}_\alpha{ k}_\beta$ 
may be replaced by $(k^2/3)\delta_{\alpha\beta}$ 
from  the angle integration of  ${\bi k}$,   
leading to   the fourth term in Eq.(2.7).

\vskip5mm
\hspace{-0.3cm}
\noindent 
{\bf Appendix B: Fixed  charge boundary condition}\\
\setcounter{equation}{0}
\renewcommand{\theequation}{B\arabic{equation}}

As illustrated  in  Fig.1, we  may 
fix   the surface changes, 
 $Q_0$ and $Q_H$, at $z=0$ and $H$ 
\cite{Landau,OnukiNATO}, as well 
as the  potential difference $\Delta\Phi$. 
Here, we consider the fixed  
charge boundary condition, where 
we assume $Q_H=-Q_0$ without 
 ionization on the surfaces. 
 See  Appendix D for the continuum theory for 
 these  two boundary conditions. 

For charged particles, let $U_{\rm e}= 
U_{\rm e}({\bi r}_1, \cdots, Q_0)$ be the  electrostatic energy 
appropriate for the fixed-charge condition, which includes 
 the contribution from the excess electrons 
on the metal surfaces. For infinitesimal changes 
${\bi r}_i \to {\bi r}_i+ d{\bi r}_i$ and $Q_0 \to Q_0+dQ_0$, 
$U_{\rm e}$  is  changed   as 
\be 
d U_{\rm e}= -\sum_i q_i {\bi E}_i \cdot d{\bi r}_i 
+ HE_a dQ_0, 
\en 
where   $E_a$ is  a dynamic variable at fixed $Q_0$. 
   The expressions for ${\bi E}_i$ 
are  the same in  the two cases at fixed $E_a$ and at fixed $Q_0$.  
  Then, from Eqs.(2.24) and (B1),  
 $U_{\rm e}$ and $U_{\rm m}$ are related  by 
\be 
U_{\rm e} = U_{\rm m}  - HL^2 E_a^2/8\pi +HE_a Q_0. 
\en

For dipoles, 
the  electrostatic energy  $U_{\rm e}^{\rm d}$ 
under the fixed-charge condition satisfies 
\be 
d U_{\rm e}^{\rm d} 
= -\sum_i  ({\bi F}_i^e \cdot d{\bi r}_i  + 
  {\bi E}_i \cdot d{\bi \mu}_i)+HE_a   dQ_0, 
\en 
which should be compared with  Eq.(3.10).
As in Eq.(B2), 
$U_{\rm e}^{\rm d}$ and $ U_{\rm m}^{\rm d}$ 
are related by   
\be 
U_{\rm e}^{\rm d} = U_{\rm m}^{\rm d}  - HL^2 E_a^2/8\pi +HE_a Q_0.
\en

\vskip5mm
\hspace{-0.3cm}
\noindent 
{\bf Appendix C: Derivation of  Eq.(2.28)}\\
\setcounter{equation}{0}
\renewcommand{\theequation}{C\arabic{equation}}

In  Eq.(2.26), $K_0(z,z')$ is given by the lateral integral,      
\bea 
&&\hspace{-9mm} K_0(z,z') = \frac{1}{4\pi } 
\int dxdy\sum_{\bi m} 
 [ \psi_\ell({|{\bi r} -{\bi r}' + 
{\bi h}|}) \nonumber\\
&&\hspace{2cm}  -\psi_\ell({|{{\bi r}} -{\bar{\bi r}}' 
 + {\bi h}|})], 
\ena 
where ${\bi r}=(x,y,z)$,  ${{\bi r}}'=(x',y',z')$, 
 ${\bar{\bi r}}'=(x',y',-z')$, and 
$0<x,y<L$.After the integration, 
the right hand side  becomes a function of 
$z$ and $z'$ independent of $x'$ and $y'$. 
We then  twice differentiate $K_0(z,z')$ with respect to $z$ and 
use  the relation  $\nabla^2 \psi_\ell(r)= -4\pi \varphi(x) 
\varphi(y)\varphi(z)$, where  $\varphi(x)$ is defined by Eq.(2.5). 
Some calculations yield   
\be  
\frac{\p^2}{\p z^2}K_0(z,z')  
= -\hat{\varphi}(z-z')+\hat{\varphi}(z+z'),
\en
where  $\hat{\varphi}(z)$ is defined by Eq.(2.29). 
The above relation is integrated to give Eq.(2.28) under 
$K_0(0,z')= K_0(H,z')=0$.  
Furthermore,  Eqs.(2.19) and (2.28) give   
\bea
&&\hspace{-1cm}\frac{\p K_0(z,z')}{\p z'} = \int_0^z du [
\hat{\varphi}(u+z')+\hat{\varphi}(u-z')]-\frac{z}{H}, \\
&&\hspace{-1cm} \frac{\p^2 K_0(z,z')}{\p z\p z'} = 
\hat{\varphi}(z+z')+\hat{\varphi}(z-z')-\frac{1}{H},
\ena 
where we have  used   
 $\int_0^H du[\hat{\varphi}(u+z')+\hat{\varphi}(u-z')]=1$ from 
the periodicity of $\hat{\varphi}(u)$. 

\vskip5mm
\hspace{-0.3cm}
\noindent 
{\bf Appendix D: Continuum theory of electrostatics}\\
\setcounter{equation}{0}
\renewcommand{\theequation}{D\arabic{equation}}

We  compare the results 
in the text and those of   continuum electrostatics 
\cite{Landau,OnukiNATO}.  
We consider charged particles in a polar medium  
between  parallel metallic plates 
under applied electric field $E_a$.  
 The system  is in the region $0<x,y<L$ and $0<z<H$. 
We assume  $L\gg H$ to  neglect the edge effect. 
We do not assume the (artificial) 
 periodic boundary condition in the $x$ 
and $y$ axes.

In this appendix,  the physical quantities 
are smooth functions of space  after 
spatial coarse-graining.  
In addition to 
the electrostatic potential $\Phi({\bi r})=\phi({\bi r})-E_az$, 
we introduce the charge density $\rho({\bi r})$ and the polarization 
${\bi p}({\bi r})$. The electric field 
${\bi E}= -\nabla\Phi=E_a {\bi e}_z -\nabla\phi  $ and the electric induction 
${\bi D}= {\bi E}+4\pi {\bi p}$ are     defined. 
For simplicity, we assume the overall  charge neutrality condition 
$\int d{\bi r}\rho=0$ without ionization on the walls. 
 Hereafter,  the integral $\int d{\bi r}$ is performed within the cell. 

From the relation $\nabla\cdot{\bi D}=4\pi\rho$, we  may 
define the effective charge density by  
\be 
\rho_e= \rho-\nabla\cdot{\bi p}, 
\en 
which satisfies  $\nabla\cdot{\bi E}=-\nabla^2\phi= 4\pi\rho_e$. 
Let $\rho_{e{\bi k}}(z)= \int d{\bi r}_\perp e^{-{\rm i}{\bi k}
\cdot{\bi r}_\perp}\rho_e$ be the 2D Fourier transformation in the $xy$ plane, 
where ${\bi k}= (k_x,k_y)$ and ${\bi r}_\perp=(x,y)$.   
 As  in Eq.(2.16), the excess potential $\phi$ is written as 
\be 
\phi({\bi r}) =4\pi\int_{\bi k}\int_0^H  dz'
G_k(z,z')\rho_{e{\bi k}}(z')  e^{{\rm i}{\bi k}\cdot{\bi r}_\perp},  
\en 
where $\int_{\bi k}= (2\pi)^{-2}\int dk_x dk_y$ 
and   the Green function $G_k(z,z')$ is given in  Eq.(2.18).

First, we  consider the lateral averages ($0<x,y<L$),
\bea   
&&\hspace{-8mm}
\bar{\phi}(z)= \int dxdy~ \phi/L^2,
\quad  
\bar{p}_z(z)= \int dxdy~ p_z/L^2,
\nonumber\\ 
&&\hspace{-9mm} \bar{\rho}(z)= \int dxdy~ \rho/L^2,   
\quad {\bar{\rho}}_e(z)= \int dxdy~ \rho_e/L^2.
\ena 
Then, we find  $\bar{\rho}_e= {\bar \rho}- d{\bar p}_z/dz$ and Eq.(D2) becomes    \be 
{\bar \phi}(z)= 4\pi \int_0^H du G_0(x,u) {\bar\rho}_e(u)  ,
\en 
where  $G_0(z,z')$ is given in Eq.(2.19). The average electric field 
$\bar{E}_z(z) =E_a -d\bar{\phi}(z)/dz$ is calculated as 
\be
{ {\bar{E}}_z(z)}=- {4\pi}  {\bar p}_z(z) + 4\pi Q_0/L^2+4\pi\int_0^z 
  du   {\bar\rho}(u) .  
\en 
The total surface charge at $z=0$ is denoted by     $Q_0$; then,  that 
at $z=H$  is  $Q_H=-Q_0$.    From Eq.(D4) we obtain \cite{commentS}    
\be 
 \hspace{-1cm}Q_0=
 \frac{L^2}{4\pi} E_a +  \frac{1}{H} 
\int d{\bi r} [z{\rho}({\bi r}) + { p}_z({\bi r})]. 
\en
This formula  corresponds to Eqs.(2.31) and (3.17).
The above relation itself 
readily follows if we set $\rho= \nabla\cdot{\bi D}/4\pi$ 
in the integral $\int d{\bi r}z\rho$.

Second, we consider the electrostatic energy 
$U_{\rm m}$ in  the fixed-potential condition. 
Its discrete versions are 
in  Eqs.(2.23) and (3.9).  The continuum version reads 
\be 
U_{\rm m}=  \int d{\bi r} 
[\frac{1}{2} \rho_e \phi -E_a(z\rho + p_z)].
\en 
For small incremental changes  ($\rho \to 
\rho+\delta\rho,  {\bi p} \to 
{\bi p}+\delta{\bi p}, ....)$,  
 $U_{\rm m}$ in Eq.(D7) is changed as  
\be 
\delta U_{\rm m}= 
\int d{\bi r}[ \Phi \delta\rho-{\bi E}\cdot\delta {\bi p}
-(z\rho +p_z) \delta E_a],
\en 
which corresponds to Eqs.(2.24) and (3.10). 
On the other hand, the electrostatic energy $U_{\rm e}$ in the fixed-charge 
condition  should satisfy 
\be 
\delta U_{\rm e}= \int d{\bi r}( \Phi \delta\rho-{\bi E}\cdot\delta {\bi p})  
+ H E_a \delta Q_0, 
\en 
which is the counterpart of Eqs.(B1) and (B3).
Then, $U_{\rm e}$ and $U_{\rm m}$ 
are related by  Eq.(B2) or (B4), leading to\cite{OnukiNATO,Landau} 
\be 
U_{\rm e}= U_{\rm m}-  \frac{HL^2}{8\pi}E_a^2 + HE_aQ_0
= \int d{\bi r} \frac{|{\bi E}|^2}{8\pi}.
\en 

Third, we remark on  the polarization ${\bi p}$. 
So far it has been treated as an independent variable. Without ferroelectric 
order,  ${\bi p}$ is usually    related to $\bi E$   by \cite{Landau},  
\be 
{\bi p}=\chi {\bi E},
\en 
in the linear response regime. 
From  ${\bi D}= \varepsilon {\bi E}$, the electric susceptibility 
$\chi$ and the dielectric constant 
$\varepsilon$ are related by 
$
\varepsilon=1+4\pi\chi.
$ 
In this situation,  we may 
introduce the following free energy contribution, 
\be 
F_{\rm p}  =\int d{\bi r}\frac{1}{2\chi} |{\bi p}|^2,  
\en  
which is an increase in the  free energy due to 
mesoscopic ordering of the constituting dipoles. 
The polarization free energy is needed to examine the thermal fluctuations 
of $\bi p$\cite{Fel,OnukiNATO}.  
We then treat  the sum,  $F_{\rm m}=U_{\rm m}+F_{\rm p} $ 
or  $F_{e}=U_{\rm e}+F_{\rm p} $, as the electrostatic free energy   
  in  the fixed-potential or fixed-charge condition. 
   Since its  functional derivative with respect to $\bi p$ 
is given by  $ -{\bi E}+\chi^{-1}{\bi p}$ 
from Eqs.(D8), (D9), and (D12), 
 its minimization yields  Eq.(D11). Eliminating $\bi p$, 
we rewrite $F_{\rm m}$ and $F_{\rm e}$  as  
\bea 
&&\hspace{-1cm}
F_{\rm m}= \int d{\bi r}(\rho\Phi - \frac{\varepsilon}{8\pi}|\bi E|^2)  
+\frac{HL^2}{8\pi}E_a^2,\\
&&\hspace{-1cm}F_{{\rm e}}=  \int d{\bi r} \frac{\varepsilon}{8\pi} |{\bi E}|^2.\ena  
The second term in $F_{\rm m}$ in Eq.(D13) 
 is a constant at fixed $E_a$, so 
 we may  redefine   the electrostatic free energy 
 as  
\be 
F_{\rm m}' = F_{\rm m}  - HL^2 E_a^2/8\pi= 
F_{\rm e}- Q_0(\p F_{\rm e}/\p Q_0),
\en 
 which is the Legendre transform of   
 $F_{\rm e} $. 
The two expressions, $F_{\rm m}'$ and $F_{\rm e}$, 
 have  both been used in the literature. 
In the Ginzburg-Landau scheme, 
Yaakov {\it et al.}\cite{Yaakov}  used $F_{\rm m}'$ 
 and one of the present authors\cite{Onuki}  
used $F_{\rm e}$ for ions  in a mixture solvent, 
where   $\varepsilon$ 
depends    on the local solvent composition and is inhomogeneous.



\begin{thebibliography}{0}
\bibitem{commentS}  
In the  Debye-H$\ddot{\rm u}$ckel theory  of electrolytes, 
the potential decays as $\exp[{-\kappa z}]$ 
or $\exp[{-\kappa (H-z)}]$ far from the walls 
for $\kappa H\gg 1$, where $\kappa$ is the Debye wave number. 
In this theory the ratio of $E_a/4\pi$ 
to $\sum_j q_j z_j /HL^2$ is $2/\kappa H$.
\bibitem{ratio} 
For the examples in Table 1, the ratio $\av{E^{{\rm d}\ell}}/E_a$ 
is given by $0.56$, 
$0.38$,$0.83$, $0.41$, $1.18$, and $0.40$  
from above.
\bibitem{Lorentz} In our simulation, 
the cut-off length for the short-range 
part of the electrostatic interaction 
is $ 4$. For $\gamma=0.65$, 
 the integral of 
$\varphi_3(r)$ in the range $r<4$ 
is nearly equal to 1.
\end{thebibliography}


\end{document}